\documentstyle[sprocl]{article}
\input{psfig}
\catcode`\@=11
\def\@oldcitex[#1]#2{\if@filesw\immediate\write\@auxout
	{\string\citation{#2}}\fi
\def\@citea{}\@oldcite{\@for\@citeb:=#2\do
	{\@citea\def\@citea{,}\@ifundefined
	{b@\@citeb}{{\bf ?}\@warning
	{Citation `\@citeb' on page \thepage \space undefined}}
	{\csname b@\@citeb\endcsname}}}{#1}}
\def\oldcite{\@ifnextchar [{\@tempswatrue
	\@oldcitex}{\@tempswafalse\@oldcitex[]}}
\def\@oldcite#1#2{{[{#1}]\if@tempswa\typeout
	{IJCGA warning: optional citation argument 
	ignored: `#2'} \fi}}
\catcode`\@=12
\newcommand{\lesssim}{\stackrel{\scriptscriptstyle<}{\scriptscriptstyle\sim}}

\def\sb{\hbox{
\begin{picture}(4,1)(0,0)
\put(0,0){SUSY}\put(0,0){\line(4,1){4}}
\end{picture}}}
 
\def\ppbar{$p\overline{p} $}            
            \def\ttbar{$t\overline{t}$}             
\def\bbbar{$b\overline{b}$}             
\def\ino{\widetilde \chi}               
\def\winog{\widetilde \chi^\pm}               
\def\zinog{\widetilde \chi^0}               
\def\winol{\widetilde \chi^\pm_1}             
\def\winoop{\widetilde \chi^+_1}              
\def\winoom{\widetilde \chi^-_1}              
\def\zinol{\widetilde \chi^0_1}               
\def\zinoh{\widetilde \chi^0_2}               
\def\w3{\widetilde W_3}               
\def\sbottom{\tilde b}             
\def\stop{\tilde t}             
\def\stau{\tilde\tau}             
\def\slepton{\tilde\ell}             
\def\selectron{\tilde e}           
\def\squark{\widetilde Q}             
\def\sneutrino{\tilde \nu}          
\def\gluino{\tilde g}             
\def\gravitino{\widetilde G}          
\def\LSP{LSP}                         
\def\tanb{$\tan{\beta}$}                           
\def\m0{$m_0$}                         
\def\mhalf{$m_{1/2}$}                     
\def\a0{$A_0$}                         
\def\pt{$p_T$}                          
\def\et{$E_T$}                          
\def\met{\mbox{${\hbox{$E$\kern-0.6em\lower-.1ex\hbox{/}}}_T$}} 
\def\sr{\hbox{\hbox{R}\kern-0.6em\lower-.1ex\hbox{/}}} 
\def\ipb{pb$^{-1}$}                     
\def\gev{GeV}                           
\def\lumint{$\int\! {\cal{L}} dt$}      
\def\D0{D\O}                            
\def \intlum {\int {\cal L} dt}
\def\goes{\rightarrow}

\def\zinoi{\widetilde \chi^0_i}               
\def\zinoj{\widetilde \chi^0_j}               

\begin{document}
\title{ The Search for Supersymmetry at the Tevatron Collider}
\author{
M. Carena,$^{1}$
R.L. Culbertson,$^{2}$
S. Eno,$^{3}$
H.J. Frisch,$^{2}$ and
S. Mrenna$^{4}$}
\address{
\centerline{$^{1}$~Fermi National Accelerator Laboratory}
\centerline{$^{2}$~University of Chicago}
\centerline{$^{3}$~University of Maryland}
\centerline{$^{4}$~Argonne National Laboratory}
}
\maketitle
\begin{abstract}
We review the status of searches for Supersymmetry at the Tevatron Collider.
After discussing the theoretical aspects relevant to the 
production and decay of supersymmetric particles at the Tevatron, we present
the current results for Runs Ia and Ib as of the summer of 1997.
\end{abstract}

\section{Introduction}
\label{sec:intro}

The Tevatron is a $p\bar{p}$ collider 
located
at the Fermi National Accelerator Laboratory (Fermilab) in Batavia, 
Illinois, USA.  Two experimental collaborations, CDF and \D0,
have collected data
at a \ppbar~center--of--mass  energy of
$\sqrt{s}=1.8$~TeV.  
The CDF detector~\cite{CDFnim} features a magnetic solenoidal 
spectrometer inside the calorimeters and a silicon vertex detector; 
the \D0 detector's~\cite{D0nim} strengths are the finely--segmented hermetic 
calorimeter and the good muon coverage.  There have been two major runs of 
the Tevatron, accumulating
approximately 20~\ipb~of data in '92--'93 and 100~\ipb~in 
'94--'95.  The next run, starting around 2000, will provide 1000~\ipb~per year.
Table \ref{tab:summary} summarizes those CDF and \D0 analyses that
have been published or presented at conferences.
This review addresses all analyses available as of summer 1997.
Due to limited space here many details must be left out; a more thorough
review is available including
lists of cuts for each analysis and
the basic theoretical tools necessary to understand the current and 
potential Tevatron analyses.\,\cite{chapter_long_version}

The Tevatron Collider has been the world's 
accelerator--based high
energy frontier since it first began taking data in 1987, and has thus been a
prime location to search for the final pieces of the 
Standard Model (SM) and new
phenomena  beyond. 
Supersymmetry (SUSY)\,\cite{SUSY}
is a new symmetry which provides a 
well--motivated extension of the SM.
If SUSY is a consistent description of Nature, then the
lower range of SUSY particle masses can be within the reach of the 
Tevatron,
motivating a wide range of searches in a large number of 
channels.\cite{tevchannels} The mass
of the lightest neutral
Higgs boson is strongly constrained within SUSY,\cite{two_loop}
and could be
within the reach of the upgraded Tevatron.\cite{higgssim,mrenna_higgs}

\section{The MSSM}
\label{section:mssm}

In the past two decades, a detailed picture of the 
Minimal Supersymmetric extension of the Standard Model (MSSM),
has emerged.\,\cite{R13A}
The details
are discussed in other chapters of this book.  In short, the MSSM contains the 
known SM particles plus SUSY partners: 
neutralinos $\zinog_{1-4}$, 
charginos $\winog_{1-2}$,
gluinos $\gluino$,
squarks $\squark_L$ and $\squark_R$,
sleptons $\slepton_L$ and $\slepton_R$,
sneutrinos $\sneutrino$, and
3 neutral Higgses $h, H$ and $A$,
plus the charged Higgs $H^\pm$.  In addition to the usual
SM parameters, the masses and interactions of the sparticles
depend on $\tan\beta$,\,\footnote{One Higgs doublet, $H_2$, couples
to $u, c,$ and $t$, while the other, $H_1$, couples to $d, s, b, e, \mu$, and $\tau$.
The parameter $\tan\beta$ is the ratio of vacuum expectation values $\langle H_2 \rangle/
\langle H_1 \rangle$ $\equiv$ $v_2/v_1$, and $v^2=v_1^2+v_2^2$, where $v$ is
the order parameter of EWSB $\simeq 246$ GeV.}
the Higgsino mass parameter $\mu$,\,\footnote{
Beware of different sign conventions
for $\mu$ and $A_f$ in the literature.  Both 
{\tt PYTHIA}\,\cite{spythia} and 
{\tt ISAJET}\,\cite{isajet} use the convention stated 
in S.P. Martin's review in this book.}  and
a number of soft SUSY--breaking ($\sb$) mass parameters which 
are added explicitly to the Lagrangian.

{\bf Sparticle Spectrum:}
The $\winog_{1-2}$ and $\zinog_{1-4}$ masses and 
their gaugino and Higgsino composition
are determined by $M_W$, $M_Z$,
$\tan \beta$, $\mu$, and
two gaugino soft $\sb$
parameters $M_1$ and $M_2$, 
all evaluated at the electroweak scale $M_{EW}$.\footnote{The electroweak
scale $M_{EW}$ is roughly the scale of the sparticle
masses themselves.  Usually, in the literature, for simplicity,
$M_{EW}\simeq M_Z$.}
The $\gluino$ mass is determined by the $SU(3)_C$ gaugino mass parameter
$M_3$.
Depending on the gaugino/Higgsinos composition,
the $\ino$ couplings to gauge bosons, and to $L$ and $R$
sfermions can differ substantially, strongly affecting production 
and decay processes.

The mass eigenstates of sfermions are, in principle,
mixtures of their $L$ and $R$ components,
and the magnitude of 
mixing is proportional to the mass of
the corresponding fermion.
The $L-R$ mixing of $\tilde u$, $\tilde d$, $\tilde c$, $\tilde s$,
$\tilde e$ and $\tilde \mu$ is thus negligible,
and the $L$ and $R$ components
are the real mass eigenstates, with
masses $m_{\squark_{L,R}}$ and $m_{\slepton_{L,R}}, m_{\sneutrino_{\ell}}$
fixed by soft $\sb$ parameters.
For $\stop$, $\sbottom$ and $\stau$, the $L-R$
mixing can be  nontrivial.  For example, the $\stop$ mixing term is
$m_t (A_t-\mu/\tan\beta)$, where $A_t$ is a 
soft $\sb$ parameter.\,$^b$
Unless there is
a cancellation between $A_t$ and $\mu/\tan\beta$, $\stop$ mixing
occurs because $m_t$ is large, and it is
possible that the lightest stop $\stop_1$ is
one of the lightest sparticles.
For the $\sbottom$ and $\stau$, $L-R$ mixing becomes important
when $m_b\tan\beta$ or $m_\tau\tan\beta$ is ${\cal O}(m_t)$,
since the $\sbottom$ mixing term, for example, is $m_b(A_b-\mu\tan\beta)$.

At tree level, the lightest CP--even Higgs boson satisfies
the relation $M_h \lesssim M_Z$, but this
result is strongly modified by radiative corrections
that depend on other MSSM parameters.\cite{Higgsrad}
The dominant radiative corrections to $M_h$ grow as
$m_t^4$ and are logarithmically dependent on the $\stop$ and $\sbottom$
masses.
Within the MSSM, a $\:${\it general}$\:$ {\bf upper}$\:$ bound 
on $\: M_h \:$ can be
determined by
a careful evaluation of the one--loop and
dominant two--loop radiative corrections.\cite{two_loop}
After setting the masses of all SUSY particles and
$M_A$ to values around 1 TeV, setting $\tan\beta>20$, and varying 
the $\stop$ mixing parameters to give the largest possible effect,
the upper bound on $M_h$ is maximized, yielding 
$M_h \lesssim 130$ GeV for $m_t=175$ GeV. 
For more moderate values of the MSSM parameters,
the upper bound on $M_h$ becomes smaller.
Given the general upper bound on $M_h$ of
about 130 GeV,
the upgraded Tevatron has the potential to provide a crucial test
of the MSSM.\cite{higgssim,mrenna_higgs}

The SUSY spectrum can be sensitive to the
exact range of $\tan \beta$.\,\cite{refB}
The measured value 
$m_t \simeq 175$ GeV defines a lower bound on $\tan \beta$ of about 1.2,
provided that the top $t$ Yukawa coupling remains finite
up to a scale  of the order of $10^{16}$ GeV.
If, instead, the $t$ Yukawa coupling should remain finite 
only up to scales 
of order of a TeV, values of $\tan \beta$    
as low as .5 would still be 
possible.\,\footnote{This implies that a perturbative
description of the MSSM would 
only be valid up to the weak scale, which  is, of course, not a very   
interesting possibility.}
Similarly,
if $\tan \beta$ becomes too large, large values of the $b$ Yukawa 
coupling are necessary  to obtain values 
of $m_b$ compatible with experiment. 
Generically, it can be shown
that values of $\tan \beta \geq 60$  are difficult to obtain if the MSSM is 
expected to remain a valid theory up to scales 
of order $10^{16}$ GeV.

{\bf Supergravity (SUGRA):} 
At present, the exact mechanism of $\sb$ is unknown. SUGRA
models assume the existence of extra superfields (the so--called 
``hidden sector'') which couple
to the MSSM particles only through gravitational--like interactions. 
These interactions are responsible for inducing the soft $\sb$ parameters.
The number of possible parameters is over 100, but
in the minimal SUGRA (mSUGRA) scenario,
all scalars (Higgs bosons, sleptons, and squarks) are assumed to have a
common squared--mass $m_0^2$, all gauginos (Bino, Wino, and gluino) have
a common mass \mhalf, and all triple--scalar couplings have the value $A_0$,
all at a scale of order $M_{\rm Planck}$ (or,
approximately, $M_{GUT}$, the scale where the gauge couplings unify).
After specifying $\tan\beta$, 
all that remains is to relate the values of the soft
$\sb$ parameters specified at $M_{GUT}$ to
their values at $M_{EW}$.  This is accomplished using
renormalization group equations (RGE's).  Moreover, $\mu$ is determined
up to a sign by demanding the correct electroweak symmetry breaking (EWSB).
Finally, the physical sparticle masses are determined as a function of
the low energy values of the soft $\sb$ parameters, which can be 
written in terms of $m_0$, $m_{1/2}$, $A_0$, $\tan\beta$, and $\mu$.\,\cite{refC}

The masses 
of the $\gluino$'s, $\winog$'s and $\zinog$'s are strongly correlated.
Once the RGE evolution is included, $\mu$
tends to be larger than $M_1$ and  $M_2$, becoming the largest as
$\tan \beta\to 1$.  
As a result, $\zinol$, $\zinoh$ and $\winol$ 
tend to be gaugino--like, and $\zinol$ can be the LSP.
The approximate mass hierarchy is
$M_{\zinoh}\simeq 2 M_{\zinol} \simeq M_{\winol} \simeq 1/3 M_{\gluino}
\simeq 0.8 m_{1/2}.$\,\footnote{In general, it may occur
that $m_{1/2}\simeq 0$. Low--energy gaugino
masses are then dominated by contributions of  
stop--top and Higgs--Higgsino loops.
In this case the $\gluino$ could be the LSP with a mass of 
a~few~GeV,
and the $\zinol$ 
may be somewhat heavier due to contributions from 
electroweak loops.\,\cite{refD}}

Because $\slepton$'s have only EW quantum numbers and the 
lepton Yukawa couplings
are small, the $\tilde e$ and $\tilde \mu$ $\sb$ mass parameters
do not evolve much from $M_{GUT}$ to $M_{EW}$ and are
roughly
given by
$m_{\slepton_L}^2 \simeq m_0^2 + 0.5 m^2_{1/2}$ and
$m_{\slepton_R}^2 \simeq m_0^2 + 0.15 m^2_{1/2}.$
The $\sneutrino$ mass is
fixed by a sum rule 
$m_{\sneutrino_{\ell}}^2 = m_{\slepton_L}^2 + M_W^2\cos 2\beta$,
and,
when \m0~is small, the $\sneutrino$ can be the LSP instead of 
the $\zinol$.  
For $\tan\beta \ge 40$, 
the $\stau$ $\sb$
mass parameters also receive non--negligible
contributions from the $\tau$ Yukawa coupling.

The $\squark$ soft $\sb$ mass parameters
evolve mainly through the strong coupling to the $\gluino$, 
so their dependence on
the common gaugino mass is stronger than for $\slepton$'s. 
For $\tilde u$, $\tilde d$, $\tilde c$ and $\tilde s$, 
the mass eigenstates are
roughly
given by
$m_{{\squark}_L}^2 \simeq  m_0^2 + 6.3 m^2_{1/2}$ and
$m_{{\squark}_R}^2 \simeq m_0^2 + 5.8 m^2_{1/2}$.
The above relations between the mass parameters 
lead to the general SUGRA prediction, $m_{\squark} \ge
0.85 M_{\gluino}$.
In general, the $\squark$'s are heavier than
the $\slepton$'s or the lightest $\zinog$ and $\winog$.
For the $L$ and $R$ components and the $L-R$ mixing
of $\stop$ and $\sbottom$ and for the Higgs soft $\sb$
parameters, the large
$t$ Yukawa coupling (and possibly the $b$ Yukawa coupling
for large $\tan\beta$) 
plays a crucial role in the RGE evolution.
As a result, the $\stop$ and $\sbottom$ can be light.
In this case, the coefficients of $m_0$ and $m_{1/2}$ have
an important dependence on
the low--energy values of the $b$ and $t$ Yukawa couplings 
which, in turn, depend on $\tan\beta$.\,\cite{chapter_long_version}
The coefficients of
\mhalf~depend on the exact values of $\alpha_s$ and the scale of
the sparticle masses.

Although mSUGRA is
defined in terms of only 5 parameters at a high scale
(\m0, \mhalf, \a0, \tanb, and 
the sign of $\mu$), it is natural to question exact universality of 
the soft $\sb$ parameters.\,\cite{nonuniv}
For example, in
a $SU(5)$ SUSY GUT model, the $\tilde e_L$ and $\tilde d_R$
reside in 
the same 5--multiplet of $SU(5)$, and may naturally have the common mass
parameter $m^{(5)}_0$ at the GUT scale.  Similarly, $\tilde u_L, \tilde d_L$, 
$\tilde u_R$, and $\selectron_R$, which reside in the same
10--multiplet, may have a common mass $m^{(10)}_0$.  
The two Higgs bosons doublets reside
in different 5-- and $\bar 5$--multiplets, with masses $m^{(5')}_0$
and $m^{(\bar 5')}_0$.  There is no known symmetry principle that
demands that all these mass parameters should be the same.  

{\bf Gauge--Mediated Supersymmetry Breaking:}
In contrast to SUGRA, the soft $\sb$ terms in the MSSM Lagrangian,
can be generated through gauge interactions at a scale
much lower than $M_{\rm Planck}$, introducing many interesting
features.  
In most models of gauge--mediated, low--energy \sb,
the gaugino and scalar masses are roughly of the same order of magnitude.
Even after RGE evolution,
sfermions with the same quantum numbers acquire the same masses
(ignoring the effects
of Yukawa couplings),
yielding a
natural mass hierarchy between weakly and strongly interacting
sfermions; the mass
hierarchy of the gauginos is fixed by the gauge couplings (as in SUGRA models).
One distinctive feature of these models
is that the spin--3/2 superpartner
of the graviton, the gravitino $\gravitino$, can be very light and
become the LSP,\,\footnote{It
is also possible to construct a model where the gluino is 
a stable LSP with a mass of a few tens of GeV.\,\cite{refEp}
In this case, the missing energy signal for SUSY
disappears, since a 
stable LSP gluino will form stable hadrons.}  
unlike in
SUGRA models, where the gravitino has a mass on the order
of the electroweak scale and is very weakly interacting.

{\bf R--Parity Violation (\sr):}
A multiplicative R--parity
symmetry is often assumed,
but one simple extension of the MSSM is to break it.\,\cite{refF}
Presently, neither experiment nor any theoretical argument 
demands R--parity conservation.
There are several effects on the SUSY phenomenology associated
with \sr~couplings: (1) lepton or baryon number violating processes
are allowed, including the
production of single sparticles (instead of pair production),
(2) the LSP is no longer stable, but
might decay to SM particles within a collider detector,
and 
(3) because it is unstable, 
the LSP need not be the $\zinog$ or $\sneutrino$, but
can be charged or colored.  Although
very strong bounds on \sr~operators can be derived from 
the present data,\,\cite{rplimits}
but there is still room for study.

{\bf Run Ia Parameter Sets (RIPS):}
Some CDF and \D0~SUSY searches 
are analyzed in the framework of so--called ``SUGRA--inspired
models.''  These RIPS are
specified by $M_{\gluino}, m_{\squark},
M_A, \tan\beta$ and the magnitude and sign of $\mu$.  
$M_{\gluino}$ defines $M_1$ and $M_2$ using 
SUGRA unification 
relations.  The
$\winog$ and $\zinog$ properties are then fixed by $\tan\beta$
and $\mu$.  In practice, the value of $\mu$ is set much larger than 
$M_1$ and $M_2$, 
so the properties of the $\zinog$'s, $\winog$'s, and $\gluino$ are similar to 
those in a pure SUGRA model. 
The first 5 flavors of $\squark$'s are degenerate in mass, with
the value $m_{\squark}$, while
the stop masses are 
$m_{\stop_1}=m_{\stop_2}=\sqrt{m_{\squark}^2+m_t^2}$, assuming also
the absence of $L-R$ mixing.
This may be quite unrealistic, since the
$\sbottom$ and $\stop$ mass can be naturally lighter.
When $m_{\squark} > M_{\gluino}$, 
the $\slepton$ masses can be fixed using approximate
SUGRA relations, and the RIPS has many features of a SUGRA model.
The region $m_{\squark} < M_{\gluino}$
is very hard to realize in SUGRA models,
but is also worth investigating.
In this case, for some analyses, a constant value of 350 GeV is used
for
$m_{\slepton_L}$, $m_{\slepton_R}$, and $m_{\sneutrino}$.
Finally, the Higgs mass $M_A$ is used to determine the Higgs boson
sector.  
In practice,
$M_A$ is set to a large value, so that the
lightest neutral Higgs boson $h$ has SM--like couplings to 
gauge bosons and fermions, and
all other Higgs bosons are heavy.

\section{ The Present Status of Sparticle Searches }

\begin{table}[!htb]
\centering
\caption{A compilation of results from Run I Tevatron SUSY searches
as of the summer of 1997.
The symbol $b$ denotes an additional $b$--tagged jet. Also listed are the
references and the section of this chapter where each analysis is
discussed. More information is available for \D0~at 
$~~http://www-d0.fnal.gov/public/new/new\_public.html$, and for
CDF at $~~http://www-cdf.fnal.gov/$}
\begin{tabular}{|l|l|l|l|l|c|c|}
\hline \hline 
Sparticle & Signature & Expt. & Run & $\intlum$(pb$^{-1})$  & Ref. &
Sec. \\ \hline
Charginos   & \met+trilepton & CDF & Ia  &  19 & \oldcite{CDF_trilepton_1a}
& 3.1 \\ \cline{2-7} 
 and& \met+trilepton & CDF & Iab & 107 & \oldcite{CDF_trilepton_1ab} & " \\ \cline{2-7}
Neutralinos         & \met+trilepton & \D0  & Ia  & 12.5 
&\oldcite{d0_wino1a} & " \\  \cline{2-7} 
& \met+trilepton & \D0  & Ib  &  95   & \oldcite{d0_wino1b} & " \\  \cline{2-7}
& $\gamma\gamma+$\met or jets & CDF & Ib & 85 & \oldcite{CDF_stop_rlc} 
                             & 4.7 \\ \cline{2-7} 
& $\gamma\gamma+$\met  & \D0 & Iab & 106 & \oldcite{d0_diphoton} & " \\ \cline{2-7} 
\hline
Squarks & \met$+\ge$3,4 jets  & CDF & Ia & 19   & \oldcite{CDF_metjets_1a}
& 3.2 \\ \cline{2-7} 
and     & \met$+\ge$3,4 jets    & \D0  & Ia & 13.5 & \oldcite{D0_metjets_1a}  & " \\ \cline{2-7} 
Gluinos & \met$+\ge$3 jets    & \D0  & Ib & 79.2 & \oldcite{D0_metjets_1b} & " \\ \cline{2-7} 
        & dilepton$+\ge$2 jets & CDF & Ia & 19   & \oldcite{CDF_metdilepton_1a} & " \\ \cline{2-7} 
        & \met$+$dilepton$+\ge$2 jets  & CDF & Ib & 81   & \oldcite{CDF_metdilepton_1b} & " \\ \cline{2-7} 
        & \met$+$dilepton     & \D0  & Ib & 92.9   & \oldcite{D0_metdilepton_1b} & " \\ \hline
Stop    & \met+$\ell+\ge$2 jets+$b$    & CDF & Ib & 90 &
\oldcite{CDF_stop_gold} & " \\ \cline{2-7}
        & \met+$\ell+\ge$3 jets+$b$    & CDF & Iab & 110 & \oldcite{CDF_stop_carmine} & " \\ \cline{2-7}
        & dilepton+jets     & \D0  & Ib & 74.5 & \oldcite{D0_stopee} & " \\ \cline{2-7}
        & \met+2 jets        & \D0  & Ia & 7.4 &  \oldcite{D0_stopjj}& " \\ \cline{2-7}
        & \met+$\gamma$+$b$  & CDF & Ib & 85 & \oldcite{CDF_stop_rlc} & 4.11 \\ \hline
Sleptons  &$  \gamma\gamma$ \met& \D0~&Iab & 106 & \oldcite{d0_diphoton} & 
3.7\\ \cline{2-7}\hline
Charged & dilepton + \met & CDF & Ia & 19 
&\oldcite{CDF_ch_higgs_jinsong}  & 3.4 \\ \cline{2-7} 
Higgs & $\tau$+2 jets+\met & CDF & Ia & 19 & \oldcite{CDF_ch_higgs_couyoumt}  & " \\  \cline{2-7} 
& $\tau$+$b$+\met+($\ell$,$\tau$,jet) & CDF & Iab & 91
&\oldcite{CDF_ch_higgs_rutgers}  & " \\ \cline{2-7} 
& $\tau$+$b$+\met+($\ell$,$\tau$,jet) & \D0 & Iab & 125
&\oldcite{D0_chhiggs}  & " \\ \hline
Neutral & $WH \goes \ell+\met$+$b$+jet & CDF& Iab &109& 
\oldcite{CDF_neu_higgs_weiming} & 3.5 \\ \cline{2-7} 
 Higgs& $WH \goes \ell+\met$+$b$+jet & \D0& Ib &100& 
\oldcite{d0_neu_higgs} & " \\ \cline{2-7} 
 & $WH,ZH \goes \gamma\gamma$+2 jets & \D0& Ib &101.2& 
\oldcite{d0_gg_higgs} & " \\ \cline{2-7} 
 & $ZH \goes b$+jet+\met & \D0& Ib &20& 
\oldcite{d0_hedin_higgs} & " \\ \cline{2-7} 
   & $WH,ZH \goes$ 2 jets+2 $b$'s      & CDF& Ib & 91  & 
\oldcite{CDF_neu_higgs_valls} & " \\ \hline  
$R$ violating & dilepton+$\ge2$ jets & CDF & Iab & 105 &
     \oldcite{CDF_Rparity} & 3.6 \\ \hline
Charged LSP & slow, long--lived particle & CDF & Ib & 90 & 
     \oldcite{CDF_stables} & 3.6 \\ \cline{2-7} 
\hline
\end{tabular}
\label{tab:summary}
\end{table}

\subsection{Charginos and Neutralinos}
\label{sec:gauginos}
\begin{figure}[!ht]
\centerline{
\hbox{
\hbox{\psfig{figure=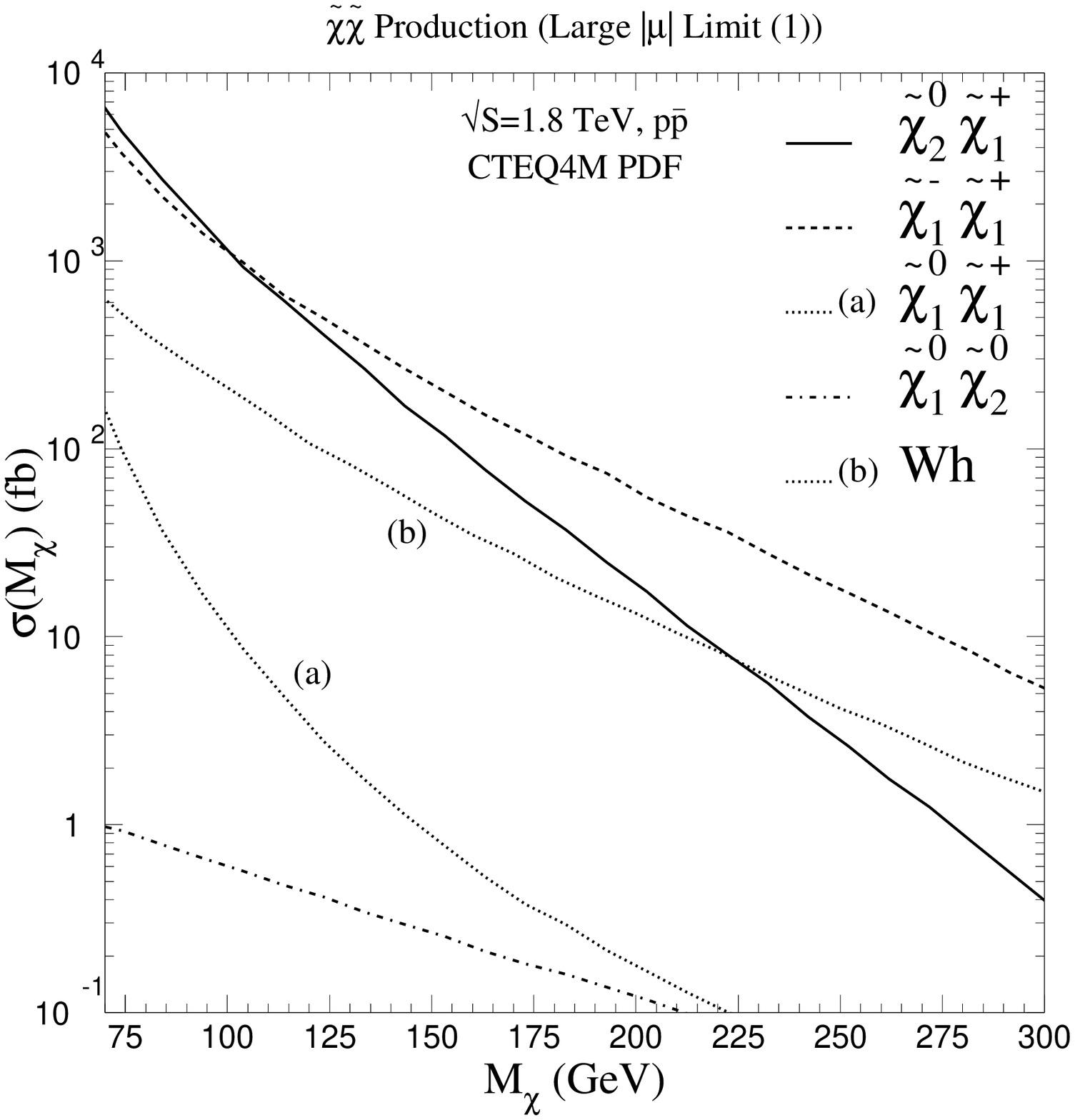,height=70mm}}
\hbox{\psfig{figure=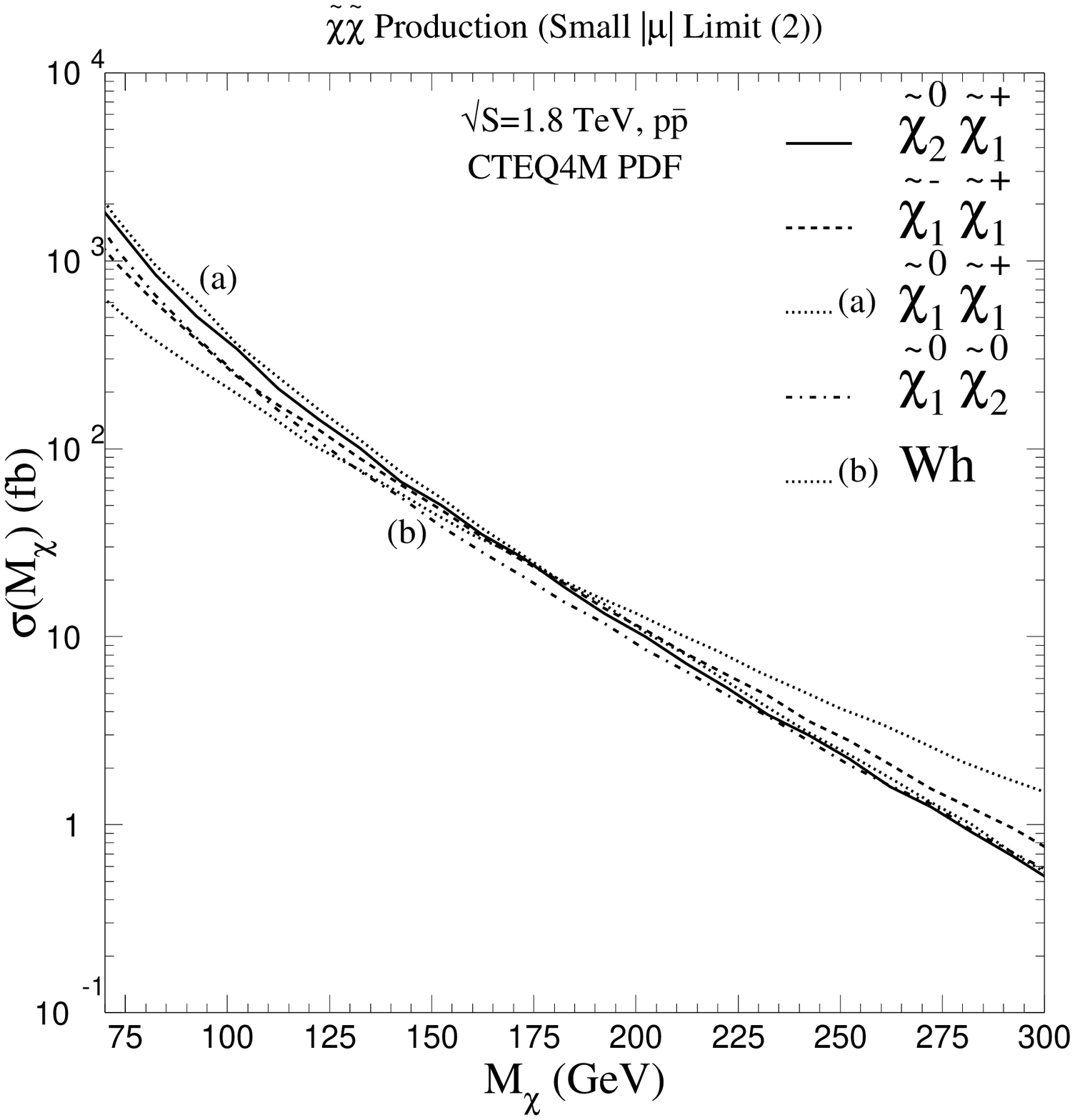,height=70mm}}
}
}
\caption{Production cross sections at the Tevatron  for $\winog$ and $\zinog$
pair production, assuming $\tan\beta=2$, $m_{\squark}=500$ GeV, and 
$M_1={1\over 2} M_2$, versus the $\winol$ mass.
The left figure is generated by fixing $\mu=-1$ TeV and
varying \mhalf, the right figure by fixing \mhalf=1 TeV and varying $\mu$. The
$Wh$ cross section (curve $b$) is shown for  reference as a function of $M_h$.
}
\label{fig:susyxsecgauge}
\end{figure}

$\winog$ and $\zinog$ pairs can be  produced  directly at hadron
colliders  through electroweak processes.
The cross sections  are 
functions of the sparticle masses and mixings, with different
contributions from t--channel $\squark$ exchange, s--channel vector
boson production, and t-- and s--channel interference.
Figure~\ref{fig:susyxsecgauge} shows the production cross sections of various
$\winog$ and $\zinog$ pairs at the Tevatron in the limits  of large
and small $|\mu|$ relative to the soft $\sb$ masses $M_1$ and $M_2$.
The decay patterns also depend on the masses and mixings.
Two--body decays dominate if allowed:
$\winog_i \to W^\pm\zinog_j$, $H^\pm\zinog_j$, 
$\winog_j h$, $\winog_j Z$,
$\tilde{\ell}_{L,R}\nu$, 
$\tilde{\nu} \ell$ or $\squark q'$ and
$\zinog_i\to Z\zinog_j$, $h\zinog_j$, $W^\mp\winog_j$, $H^\mp\winog_j$, 
$\tilde{\ell}_{L,R}\ell$, $\tilde{\nu} \nu$, or $\squark q$. 
When no 2--body final states are kinematically allowed,
3--body 
decays like $\winog_i\to\zinog_j f\bar{f}'$, 
$\winog_i\to\winog_j f\bar{f}$, $\zinog_i\to\winog_j f\bar{f}'$, 
and $\zinoi\rightarrow\zinoj f\bar{f}$ 
occur through virtual sfermions
and gauge bosons.
If the sparticles 
are light enough, 
a 100\% branching ratio to $\ell\nu$, $\ell^+\ell^-$ or $jj$ final states is 
possible.  The one--loop decay $\zinoh\to\zinol\gamma$ can 
be important if the $\zinol$~is 
Higgsino--like and $\zinoh$~is gaugino--like, or {\it vice versa}.
For a light enough $\gluino$, the decays $\winog_i\to \gluino f\bar{f}'$
and $\zinog_i\to\gluino f\bar f$ can be important.

The production of $\winog_1\zinoh$, followed by the decays $\winog_1\to\zinol
\ell\nu$ and $\zinoh\to\ell^+\ell^-\zinol$, is a source of three charged leptons
($e$ or $\mu$) and \met. This
trilepton signal has small SM backgrounds, and is consequently one of the 
``golden'' SUSY signatures.\,\cite{ref2}

\begin{figure}[!ht]
\centerline{
   \hbox{
      \hbox{\psfig{figure=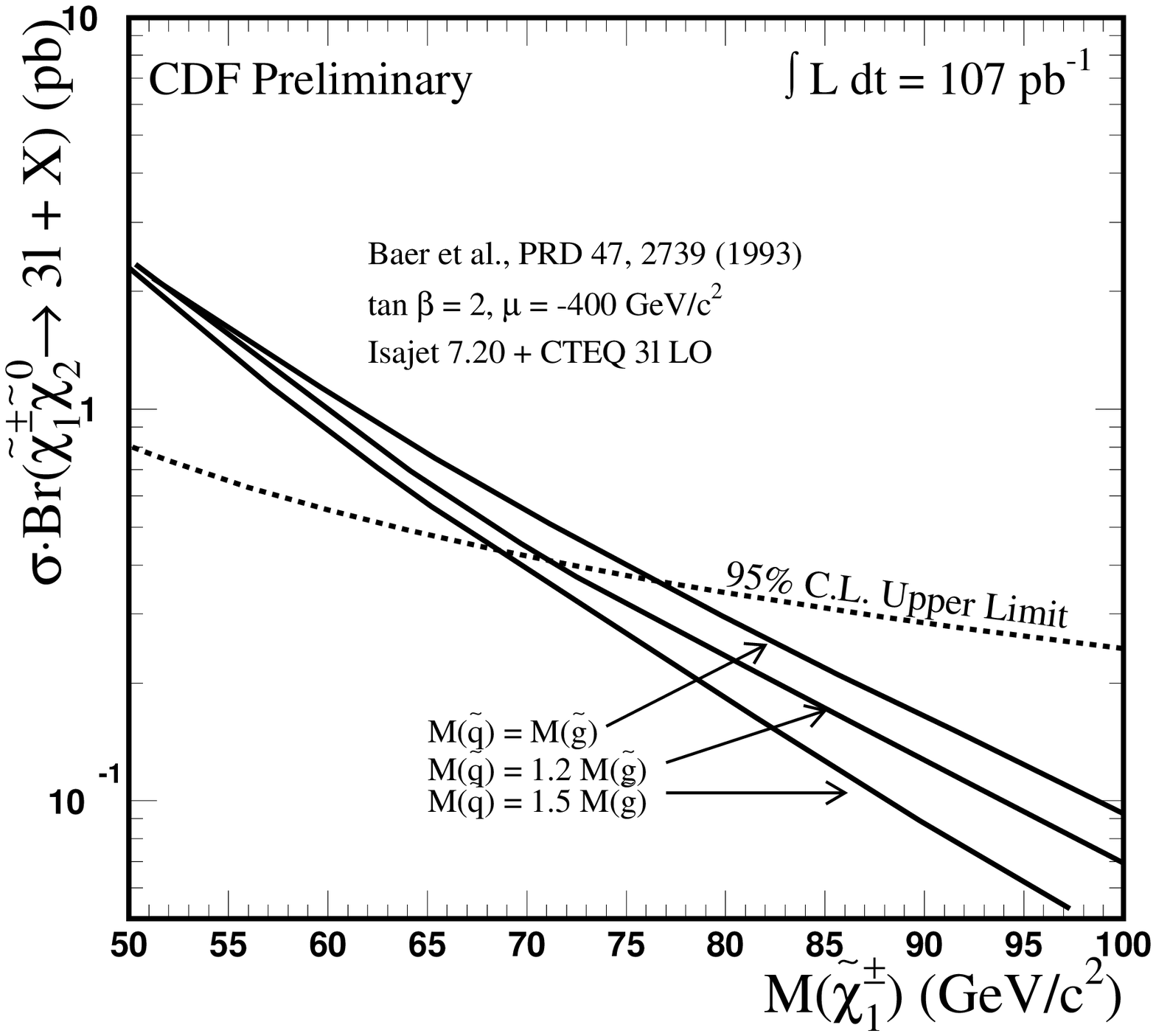,height=75mm}}
      \hbox{\psfig{figure=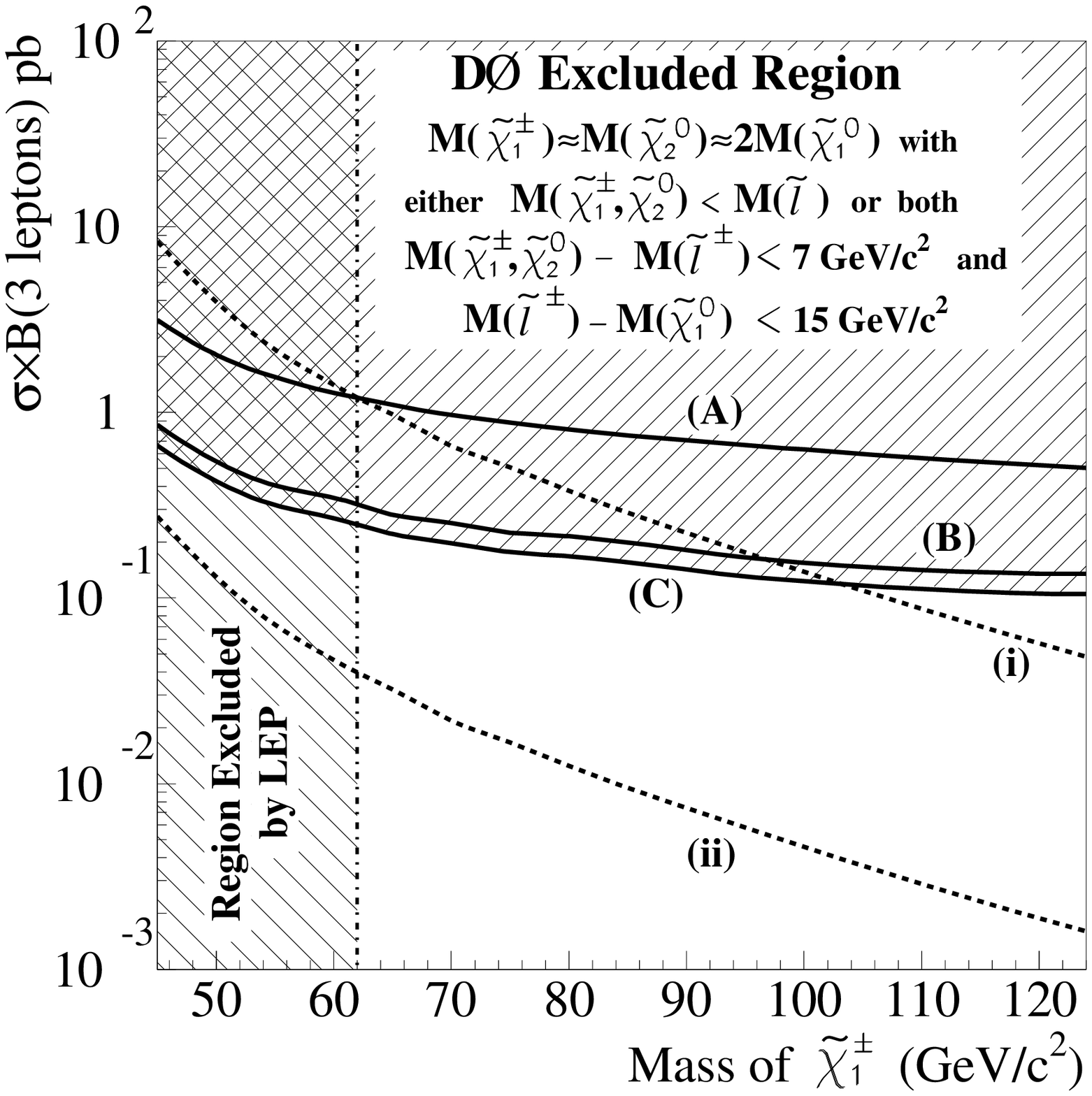,height=70mm}}
   }
}
\caption{
(Left) The CDF 95\% C.L. limits on cross section 
$\times$ branching ratio for $\winol \zinoh$ production
in 107 \ipb of data.  The limit is on the {\it sum} of the final states
$eee, ee\mu, \mu\mu\mu$ and $\mu\mu e$ when
$\winol\rightarrow \ell\nu\zinol$ and $\zinoh\rightarrow\ell\ell\zinol$.
The signals expected for three different RIPS scenarios are shown for
comparison.\protect\cite{hbaerslepton}
(Right) Similar limits from \D0, but for the {\it average} of
all four channels.
The curves (A), (B), and (C) show the Run Ia, Run Ib, and combined limits.  
Curve (i) shows the predicted cross section $\times$ branching ratio 
assuming BR($\winol\to\ell\nu\zinol$)=
BR($\zinoh\to\ell^+\ell^-\zinol$)=1/3 ($\ell=e,\mu,\tau$).
Curve (ii) assumes
BR($\winol\to\ell\nu\zinol$)=$0.1$ and 
BR($\zinoh\to\ell^+\ell^-\zinol$)=$0.033$.
For both CDF and \D0,
kinematic efficiencies are calculated using the production cross 
section from {\tt ISAJET}.
}
\label{fig:cdfwino}
\end{figure}

The results of the CDF~\cite{CDF_trilepton_1a,CDF_trilepton_1ab} and
\D0~searches~\cite{d0_wino1a,d0_wino1b}  are shown in Fig.~\ref{fig:cdfwino}
analyzed using RIPS.  The searches include 4
channels: $e^+e^-e^\pm$, $e^+e^-\mu^\pm$,  $e^\pm\mu^+\mu^-$ and
$\mu^\pm\mu^+\mu^-$.   
The CDF analysis requires one lepton with $E_T>$11~GeV, passing
tight identification cuts, and two other leptons with $E_T>$5~GeV ($e$)
or  $p_T>$4~GeV ($\mu$), passing loose identification cuts.  
All leptons must be
isolated, meaning there is little excess $E_T$ in a cone of size $R=0.4$ in
$\eta-\phi$ space centered on the lepton. 
The event must have two leptons with  the same
flavor and opposite sign.  If two leptons of the same 
flavor and opposite charge
have a mass consistent with the $J/\Psi$, $\Upsilon$ or $Z$ boson, the event is
rejected.  After this selection, 6 events remain in the data set, while the
expected background, dominated by Drell--Yan pair 
production plus a fake lepton, is
8 events. After demanding  \met$>15$~GeV, no events remain, while 1.2 are 
expected from SM model sources.

The \D0~analysis requires 3 leptons with $E_T>5$~GeV.
However, several different triggers
are used, and  some lepton categories are required to have a larger $E_T$ to
pass the various trigger thresholds.   
All leptons are required to be isolated. 
To reduce events with mismeasured \met, 
the \met~direction must not be along or opposite to a
muon. Additional cuts are tuned for each topology.  For example,  the 
background from Drell--Yan pair production plus a fake lepton 
is highest in the $eee$ channel, so these
events are rejected if an electron pair is back--to--back. 
The \met~threshold is
15~GeV for $eee$, and 10~GeV for the other three topologies. No events are
observed in any channel with a total of 1.26 events expected  from $(i)$
Drell--Yan production plus a fake lepton and $(ii)$ heavy--flavor
production.

The \D0~limit is on the ``average'' 
of the 4 modes, while the CDF limit is on the sum. After
accounting for this difference, the CDF limit is twice as
sensitive at a given $\winol$~mass.
The CDF limit shown is compared to three RIPS, which have different 
ratios of $m_{\squark}$ to $M_{\gluino}$.  The \D0~limit is compared to
a wide variation of possible branching ratios.
The \D0 theory curve assumes heavy $\squark$'s, 
which reduces the cross section, but the CDF curves do not.
The wide differences in the theory curves
in Fig.~\ref{fig:cdfwino} show the dangers of quoting a mass limit rather than
a cross section $\times$ branching ratio limit.\cite{cmssm}

\subsection{Squarks and Gluinos}
\label{sec:squark_gluinos}

\begin{figure}[ht]
\centerline{
\hbox{
\hbox{\psfig{figure=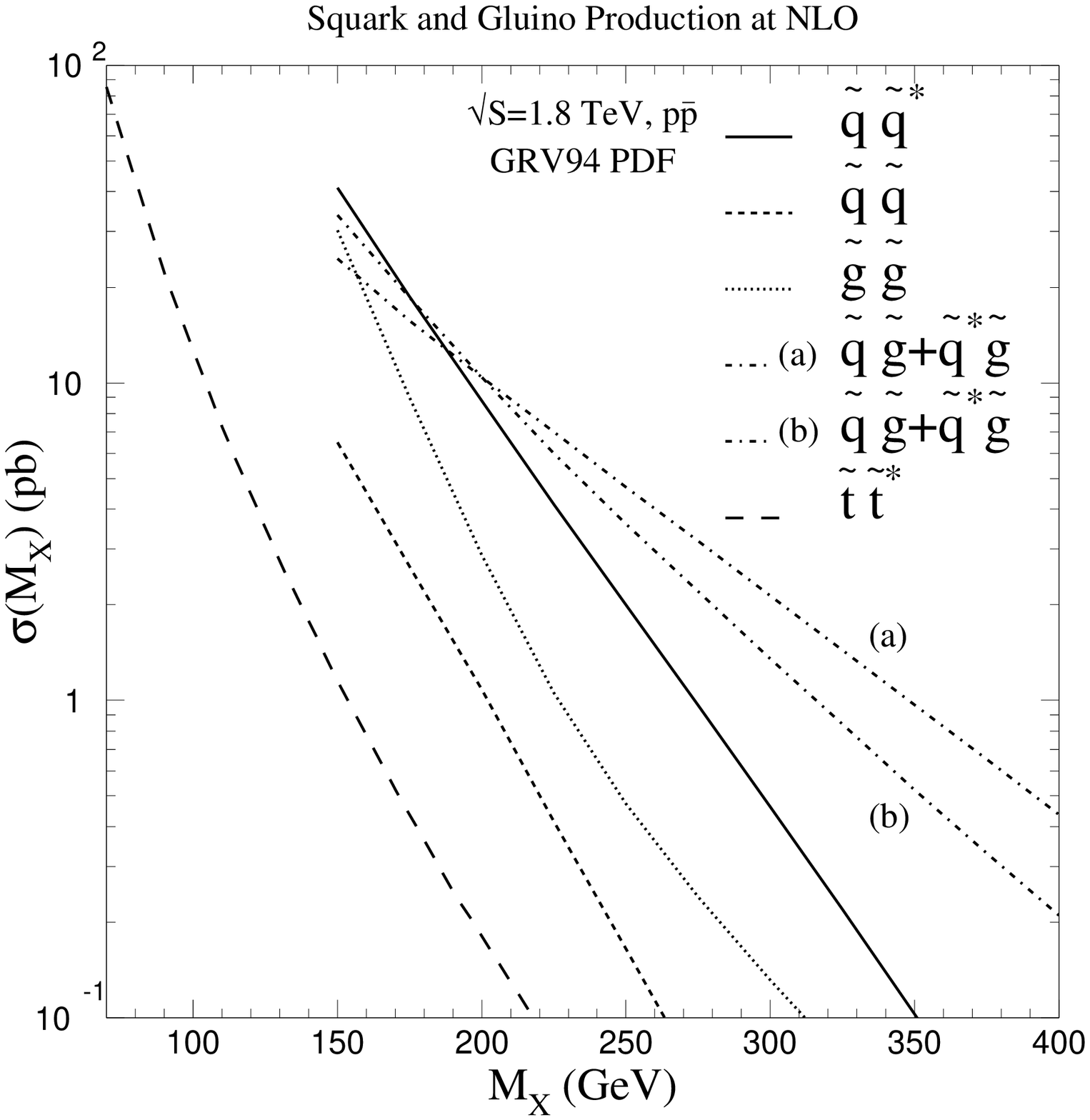,height=60mm}}
\hbox{\psfig{figure=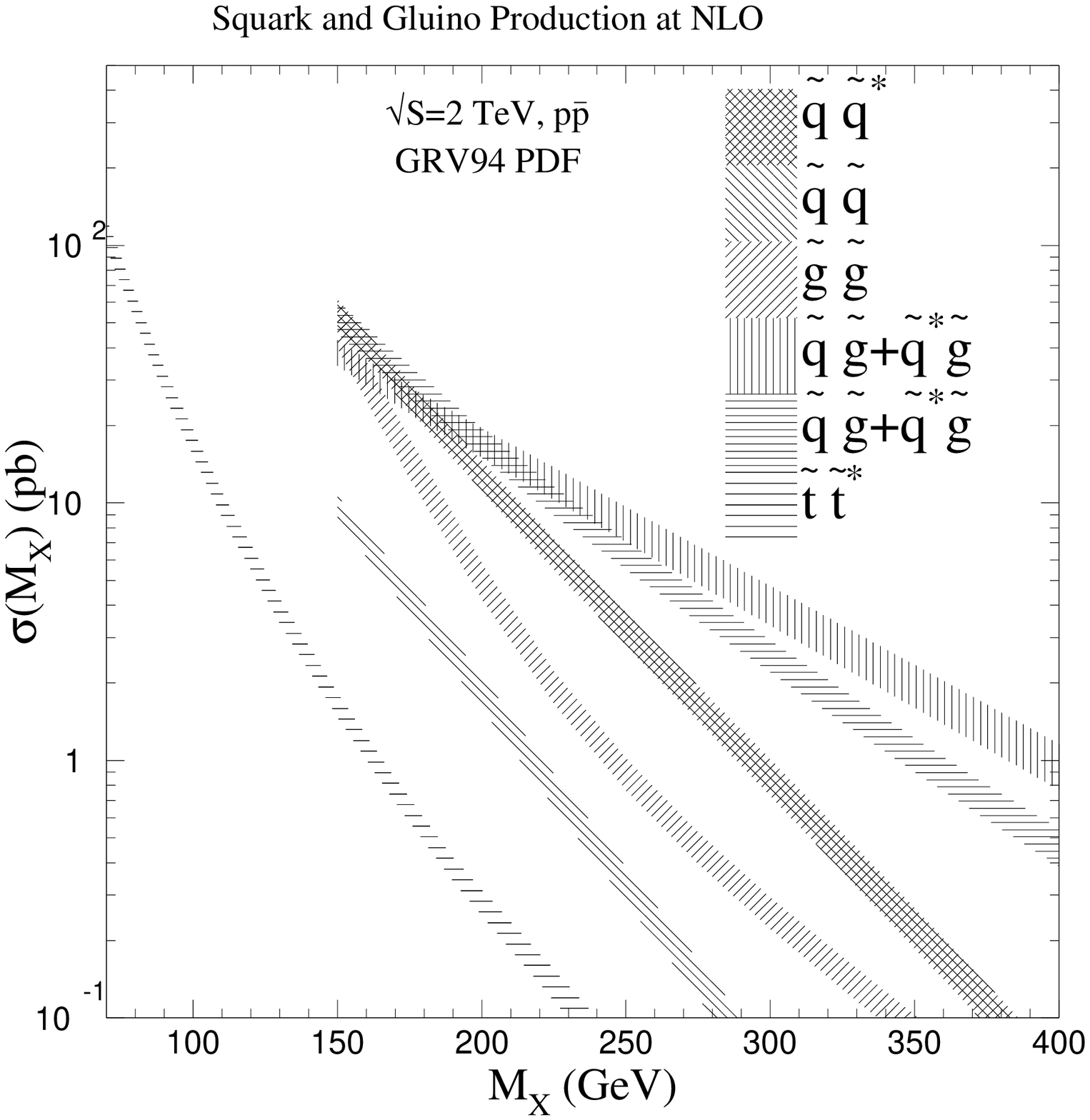,height=60mm}}}
}
\caption[0]
{(Left) Production cross sections for $\gluino$'s and $\squark$'s versus
sparticle mass $M_X$ at the Tevatron, $\sqrt{s}=1.8$ TeV, assuming
degenerate masses for 5 flavors of $\squark$'s.  For $\squark\squark^*$
and $\squark\squark$ production, $M_X$ is the squark mass and 
$M_{\gluino}=200$ GeV.  For $\gluino\gluino$ production, $M_X$ is
the $\gluino$ mass and $M_{\squark}=200$ GeV.  For $\squark\gluino$
production $(a)$, $M_X$ is the $\squark$ mass and $M_{\gluino}=200$ GeV;
for $(b)$, $M_X$ is the $\gluino$ mass and $M_{\squark}=200$ GeV.  
For $\stop\stop^*$ production, $M_X$ is the stop mass.  
All cross sections are evaluated at a scale equal
to the average mass of the two produced sparticles.
(Right) The same curves with $\sqrt{s}=2$ TeV.
The bands show the change in rate from varying the scale
from $1/2$ to $2$ times the average mass of the produced particles.}
\label{fig:susyxseccol}
\end{figure}

The Tevatron is a hadron collider, so 
it can produce $\gluino$'s and $\squark$'s through their
$SU(3)_C$ couplings to quarks and gluons.  
The  dominant production mechanisms are 
$gg,q\bar q\to \gluino\gluino$ or $\squark\squark^*$,
$qq\to \squark\squark$ and $qg\to\squark\gluino,\bar q g\to 
\squark^*\gluino$,  
and the cross
sections can be calculated
as a function of only the $\squark$ and $\gluino$ masses
(ignoring EW radiative corrections).
Figure \ref{fig:susyxseccol} shows the cross sections for
$\squark$'s and $\gluino$'s as a function of the sparticle masses 
at $\sqrt{s}=1.8$ TeV (left) and 2 TeV (right),
where NLO SUSY QCD corrections  have been included.\,\cite{squarknlo}
The NLO corrections are in general significant, positive 
(evaluated at a scale equal to the average mass of the two 
produced particles), 
and much less sensitive to the choice of scale than
a LO calculation.

Since $\squark$'s and $\gluino$'s decay into $\winog$'s and $\zinog$'s,
their signatures can be similar to $\winog$ and $\zinog$ pair production,
but with accompanying jets.
If $m_{\squark} > M_{\gluino}$, then the $\squark$ has
the 2--body decay $\squark \rightarrow \gluino q$.
The $\gluino$  has then the possible decays 
$\gluino\to q \bar{q}\zinog_i$ or
$\gluino\to q \bar{q}'\winog_i$,
 where $q$ can stand for $t$ or $b$ as well, or even
$\gluino\to t\stop^*$ or $\bar t \stop$ if kinematically
allowed.
The $\gluino$ can also decay at one--loop like 
 $\gluino\to g \zinog_i$.
If, instead, $m_{\squark} <  M_{\gluino}$, then the $\gluino$
has the 2--body decay
$\gluino \to \squark q$.
The  $\squark$'s can then decay as
$\squark_{L,R} \rightarrow q \zinog_i$,      
$\widetilde{u}_{L } \rightarrow d \ino^+_i$,  and       
$\widetilde{d}_{L } \rightarrow u \ino^-_i$.
The $\gluino$'s and $\squark$'s may also be produced in association 
with $\winog$'s or $\zinog$'s (analogous to $W+j$ and $Z+j$ production).  
Event signatures are similar to $\squark$ and $\gluino$ production,
but possibly with fewer jets. 
Promising signatures for $\squark$ and $\gluino$ production are $(i)$
multiple jets and \met\,\cite{ref1} and $(ii)$ isolated leptons and jets and 
\met.\,\cite{ref3}

{\bf Jets + \met:}
Both CDF and \D0 have performed searches for events with jets and \met.
This signature has significant physics
 and instrumental backgrounds.
The three  dominant physics backgrounds are 
$(i)$ $Z \to \nu\bar\nu$ + jets,
$(ii)$ $W\to \tau \nu$ + jets, where the $\tau$ decays hadronically, 
and $(iii)$ \ttbar$\to\tau$ + jets, where the $\tau$ decays hadronically.  
The \met~in leptonic $W$ decays
peaks at $M_W/2\simeq 40$~GeV, 
with a long tail at high \met~due to off--shell or
high--\pt~$W$'s and energy mismeasurements, so a large \met~cut is needed 
to remove these events.
Instrumental backgrounds come from mismeasured vector boson, $t$, 
and QCD multijet events.  
Backgrounds from
vector boson and $t$ production occur for  
$W \to e\nu,\mu \nu$ + jets  events
when the lepton is lost in a crack or is 
misidentified as a jet.  
QCD multijet production is a
background when jet energy mismeasurements cause false \met. 

\begin{figure}[!ht]
\center                     
\centerline{
\psfig{figure=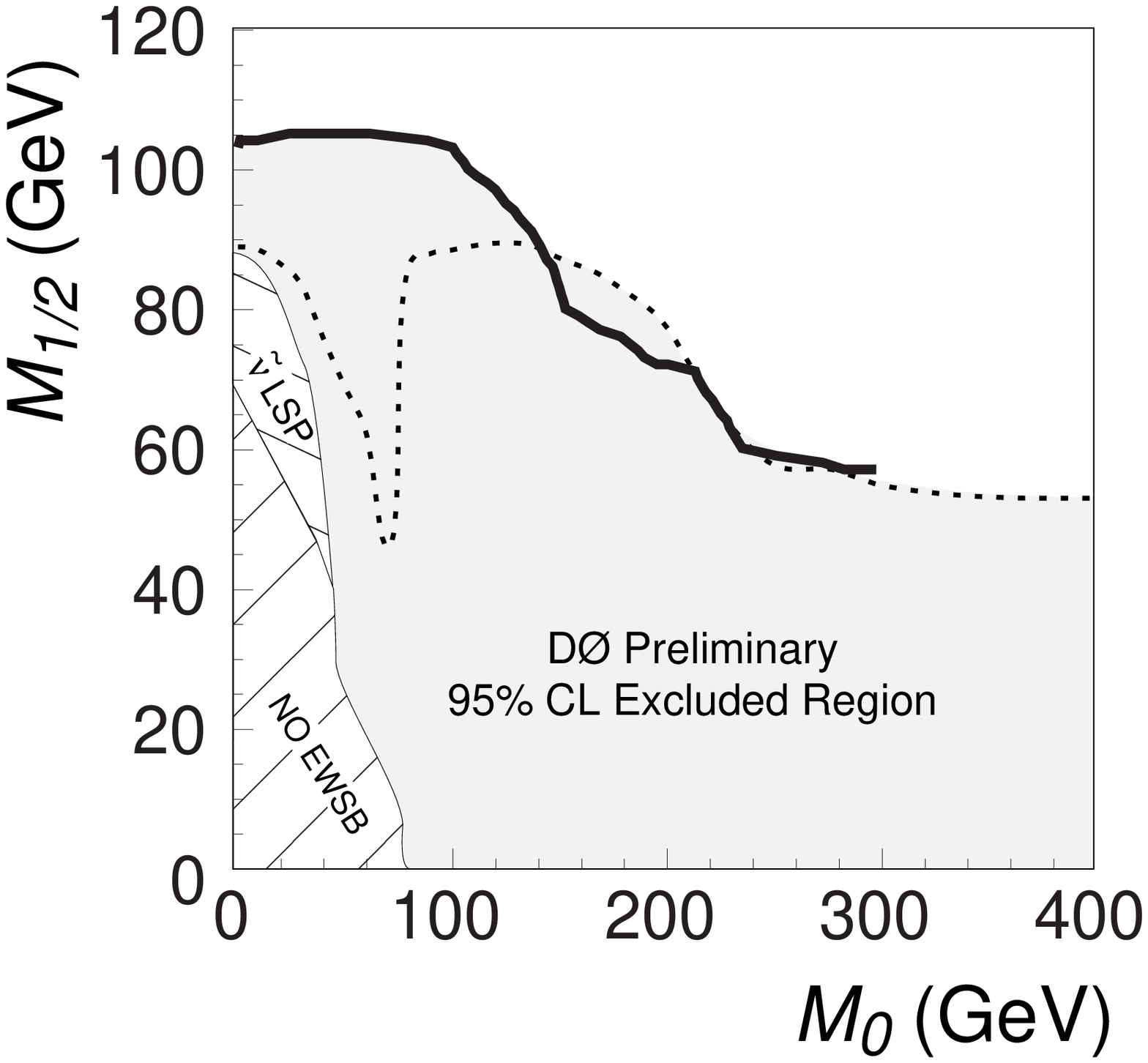,height=60mm}
\psfig{figure=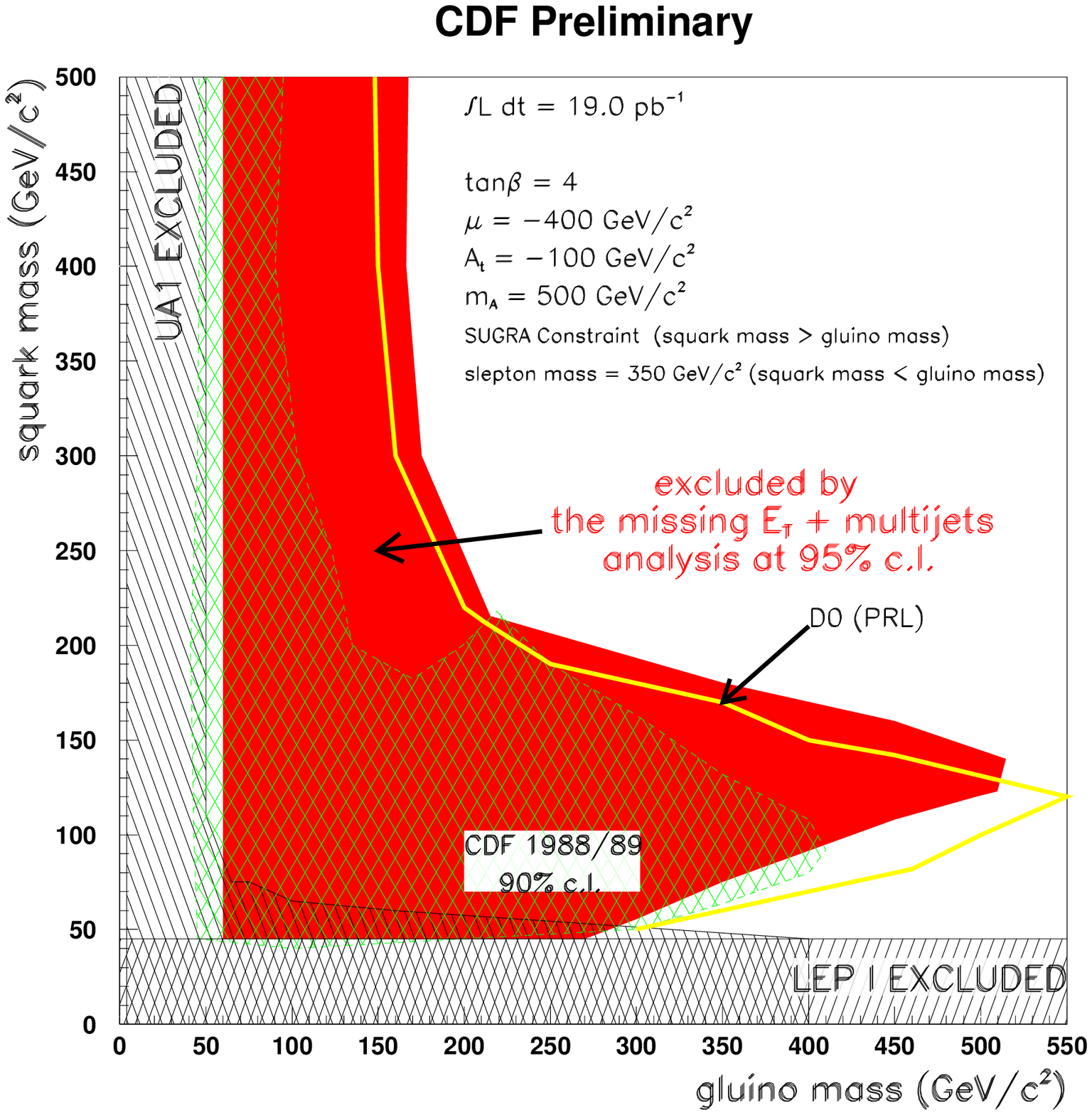,height=60mm}}
\caption{
(Left)  The \D0~Excluded region in the $m_0-m_{1/2}$ plane with fixed
parameters $\tan\beta = 2$, \a0 = 0, and $\mu<0$. The heavy solid line
is the limit contour of the \D0~jets and \met~analysis. 
The dashed line is
the limit contour of the \D0~dielectron analysis. The lower hashed 
area is a region where mSUGRA does not predict EWSB correctly.
The hashed region above is where the $\sneutrino$ is the LSP.
(Right) The CDF mass limits on $\squark$'s and $\gluino$'s from the search 
in jets and \met\,\protect\cite{CDF_metjets_1a}
using 19 \ipb~of data and the ISAJET 7.06 
Run I Parameter Set (RIPS) with the indicated values (solid area).
For $m_{\squark}<M_{\gluino}$,
the cross section used is LO, and 3 or more jets are required.
For $M_{\gluino}<m_{\squark}$, 
the cross section is NLO,\,\protect\cite{squarknlo} 
and four jets are required.
The line labelled 
``\D0 PRL'' is the \D0~result from Run Ia using 13.5 \ipb~of 
data.\,\protect\cite{D0_metjets_1a}
}
\label{fig:cdfd0squark}
\end{figure}

The \D0~Run Ia analysis\,\cite{D0_metjets_1a} 
searches for events with 3 or more jets and \met~and 
with 4 or more jets and \met.  Jets have $E_T>$20~GeV and cannot
point along the \met~to avoid backgrounds from energy mismeasurement.
The \met~threshold is 65~GeV and leptons are vetoed to remove 
$W$ backgrounds.  The resulting 
mass limits on $\squark$'s and $\gluino$'s
are shown in Fig.~\ref{fig:cdfd0squark} ((right), the plot
containing the CDF results also shows the \D0 Ia results);
these limits were
set using a RIPS model with the parameters
$M_{H^{\pm}}$= 500 GeV,
\tanb = 2, $\mu=-250$ \gev, and  
$M_{\slepton}=m_{\squark}$.\,\footnote{The efficiency and 
theoretical cross sections were calculated
using {\tt ISAJET} assuming 5 flavors
of mass degenerate $\squark$'s without $\stop$ production.}

\D0~also has a 3--jet analysis \cite{D0_metjets_1b} based on 79.2 \ipb~of  Run
Ib data.  Jets must have $E_T>$25~GeV; the leading jet must
have $E_T>$115~GeV
because the only available unbiased sample to
study the QCD multijet background had this requirement.
The \met~cut ranges from $75-100$~GeV and the 
$H_T$ (scalar sum of the non--leading jet $E_T$'s) cut ranges from 
$100-160$~GeV, optimized for each point in parameter space.
Vector boson
backgrounds are estimated  using {\tt VECBOS},\cite{vecbos} 
while the $t\bar t$
background  uses {\tt HERWIG}\,\cite{herwig} normalized to the \D0~measured
$t\bar t$ cross section.   Two techniques were used to calculate the QCD
multijet background. One compares the opening angle between the two leading 
jets
and the \met~in the signal sample to the analogous
distribution in a generic multi--jet
sample.  The other selects events from a single jet trigger which pass all the
selection criteria except for  the \met~requirement. The \met~distribution is
fit in the low \met~region, and extrapolated into the signal region. The
background estimates can be found in Table~\ref{tab:squarkevts3}.

The \D0 data have been analyzed in the context of a mSUGRA model. 
For fixed \tanb, \a0, and sign of $\mu$, exclusion curves are plotted in the
$m_0-m_{1/2}$ plane (Fig.~\ref{fig:cdfd0squark} (left)). The limits are from the
3--jet, 79.2\ipb, analysis only.$^h$ 
For each point in the  limit plane, the \met~and
$H_T$ cuts are reoptimized based on the predicted background and SUSY signal.
These results are robust within the mSUGRA framework.\cite{D0_metjets_1b} 

The CDF analysis of the Run Ib data set is not yet finished,
but the Run Ia result based on 19 \ipb~has been published.\cite{CDF_metjets_1a}
The basic requirements are 3 or 4 jets with $E_T>$15~GeV and \met$>60$~GeV.
The vector boson backgrounds are estimated using
{\tt VECBOS}
normalized to the CDF $Wjj$ data.
The $t\bar t$ backgrounds are determined using {\tt ISAJET}
normalized to 
the CDF measured cross section.
The QCD background is estimated using an independent data sample
based on a trigger that required one jet with $E_T>50$~GeV.
First all analysis cuts 
are applied to this sample 
except for the $S$ cut, the \met~cut, and the
3 or 4 jets cut.  Next the \met~distribution is fit and the number of events 
expected to pass the \met~cut is derived.  Finally the efficiency of the 
last three cuts is applied to arrive at the final background estimate,
shown in Table~\ref{tab:squarkevts3}.

The limits derived from the CDF analysis are shown in 
Fig.~\ref{fig:cdfd0squark} (right) within the RIPS framework.
In RIPS, a heavy $\gluino$ implies a heavy
$\zinol$, so a light $\squark$ ($m_{\squark} \approx M_{\zinol}$)
decay will not produce much \met.  The
consequence is an apparent hole in the CDF limit for small $m_{\squark}$ and
large $M_{\gluino}$.  However, lighter $\gluino$'s produce a large
\met~because of the enforced mass splitting between the ${\gluino}$ 
and ${\zinol}$.
The results of this analysis do not change substantially  
as parameters are varied 
within the RIPS framework.\cite{CDF_metjets_1a} 

\begin{table}[!ht]                              
\centering
\caption{The number of expected and observed 
events for Tevatron $\squark$ and $\gluino$ searches 
in the jets+\met~channels.}
\begin{tabular}{|l|c|c|c|c|}  \cline{2-5}
\multicolumn{1}{c|}{}   &  \multicolumn{2}{c|}{\D0} &
                 \multicolumn{2}{c|}{CDF} \\  \hline\hline
Analysis          &  3 jets    &  4 jets      &  3 or 4 jets      &  4 jets   \\ \hline
\lumint (\ipb)    & 79.2       &  13.5        &  19          & 19        \\ \hline
$W^{\pm}$         & $1.56\pm.67\pm.42$  & $4.2\pm1.2$  & $13.9\pm2.1\pm6.0$ & $2.6\pm0.9\pm1.7$ \\ 
$Z\to\ell\bar{\ell},\nu\bar{\nu}$ 
                  & $1.11\pm.83\pm.36$       & $1.0\pm0.4$  & $5.0\pm0.9\pm2.7$  & $0.4\pm0.2\pm0.4$ \\ 
\ttbar            & $3.11\pm.17\pm1.35$       & --           & $4.2\pm0.3\pm0.5$ & $2.2\pm0.2\pm0.4$ \\ \hline
QCD multijets     & $3.54\pm2.64$       & $1.6\pm0.9$  & $10.2\pm10.7\pm4.2$ & $3.2\pm3.8\pm1.3$ \\ \hline
Total Background  & $9.3\pm0.8\pm3.3$ & $6.8\pm2.4$ & $33.5\pm11\pm16$ & $8\pm4\pm4$     \\ \hline
Events Observed   & {\bf 15}   & {\bf 5}     &  {\bf 24}   & {\bf 6}   \\ 
\hline \hline
\end{tabular}
\label{tab:squarkevts3}
\end{table}

{\bf Dileptons+\met:}
If, in the cascade decay chain of the $\squark$'s and $\gluino$'s, 
there are two decays $\winol\to\ell\nu\zinol$, or one 
decay $\zinoh\to\ell^+\ell^-\zinol$, the final state can contain
2 leptons, jets, and \met, which is a
relatively clean experimental signature. 
The requirement of two leptons significantly reduces jet
backgrounds and removes most of the $W$ backgrounds.
Cutting on lepton pairs with the $Z$ mass
removes most of the $Z$ backgrounds.
If the leptons are required to have
\pt$>$20~GeV, the major background from
physics processes
is \ttbar$\to bW^+\bar bW^- \to b\bar b\ell^+\ell^-$\met.
As the cut on lepton \pt~is lowered,
$Z \to \tau^+ \tau^-$, where the $\tau$'s decay semileptonically,
also becomes an important background.
The instrumental backgrounds are small.   The spectacular signature
of {\it like--sign}, isolated dileptons, which is difficult to
produce in the SM, can occur 
whenever a $\gluino$ is produced directly or in a cascade decay, since
the $\gluino$ is a Majorana particle.  This property is exploited in
the CDF dilepton searches.

\begin{figure}[!ht]
\center
\centerline{\psfig{figure=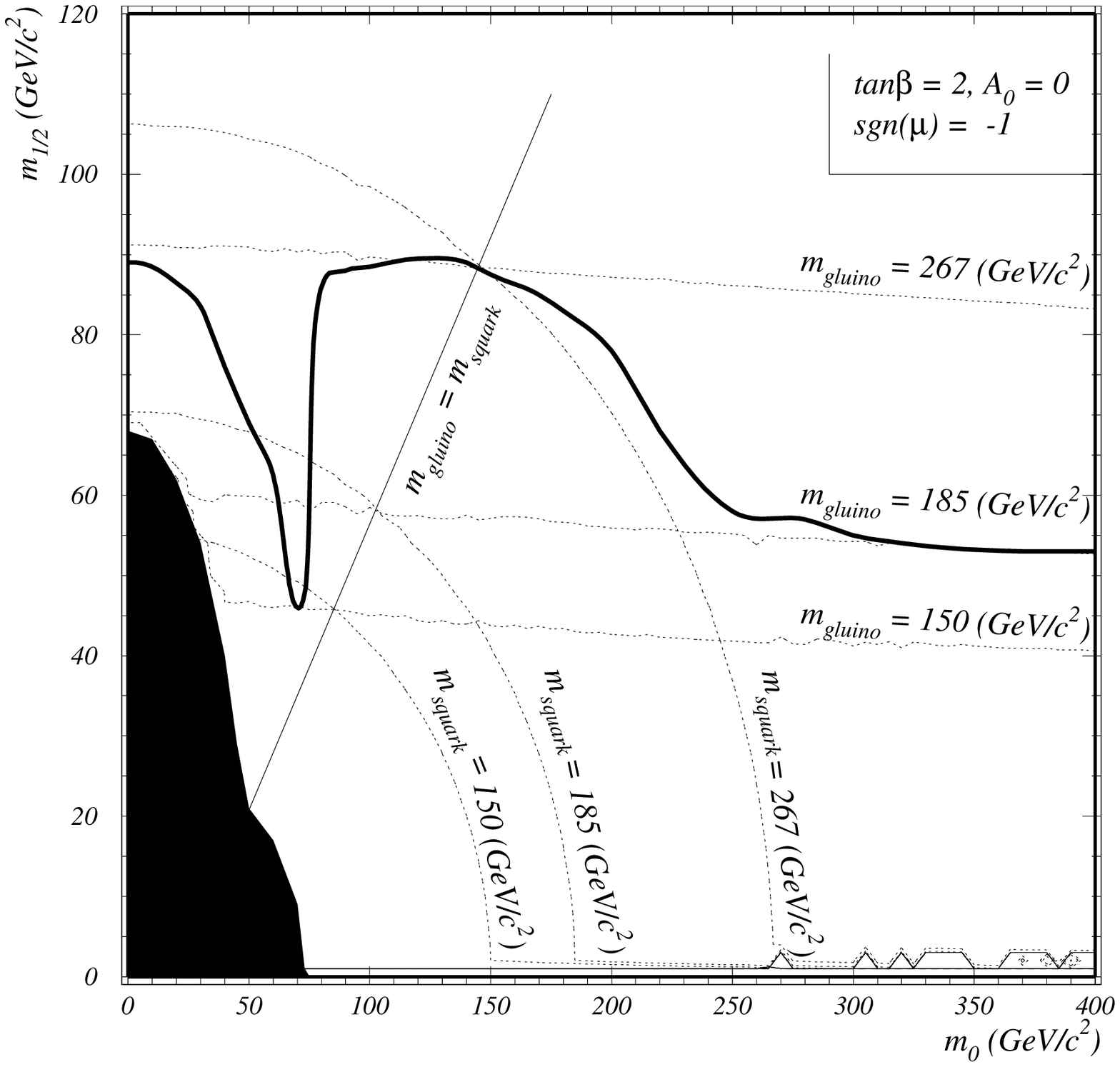,height=60mm}
\psfig{figure=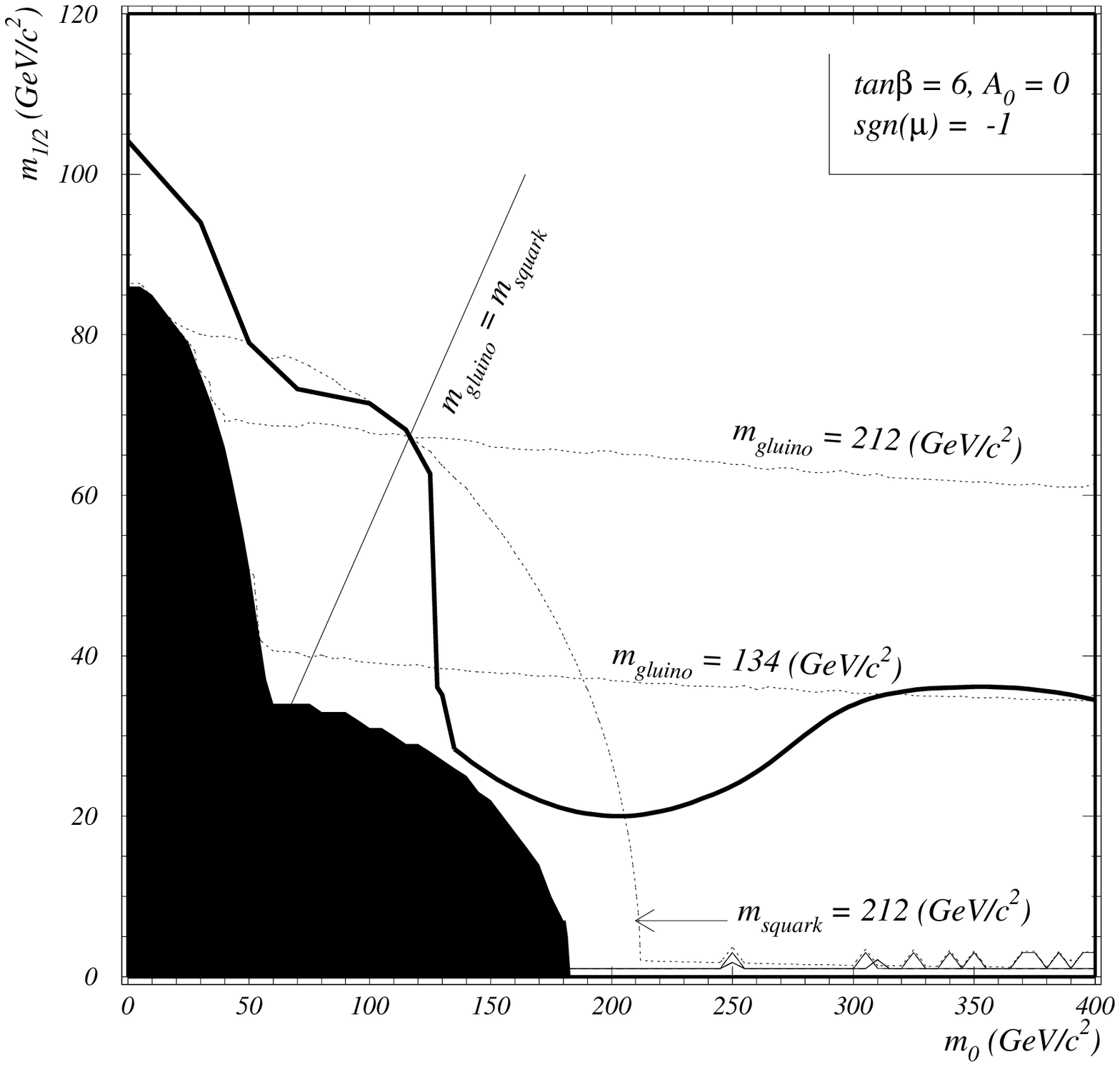,height=60mm}}
\caption{(Left) The \D0~limits on the SUGRA parameters \m0~and \mhalf~from the 
2 leptons, 2 jets, and \met~search\,\protect\cite{D0_metdilepton_1b}
for \tanb=2, \a0=0, and $\mu<0$.  (Right) The same
plot for \tanb=6, \a0=0, and $\mu<0$.  In both plots, the
dark shaded area is the region in which 
electroweak symmetry breaking is not realized.  Selected contours
of $\squark$ and $\gluino$ mass are also shown.
}
\label{fig:d0squarkee}
\end{figure}

\begin{figure}[!ht]
\center
\centerline{
\hbox{
\hbox{\psfig{figure=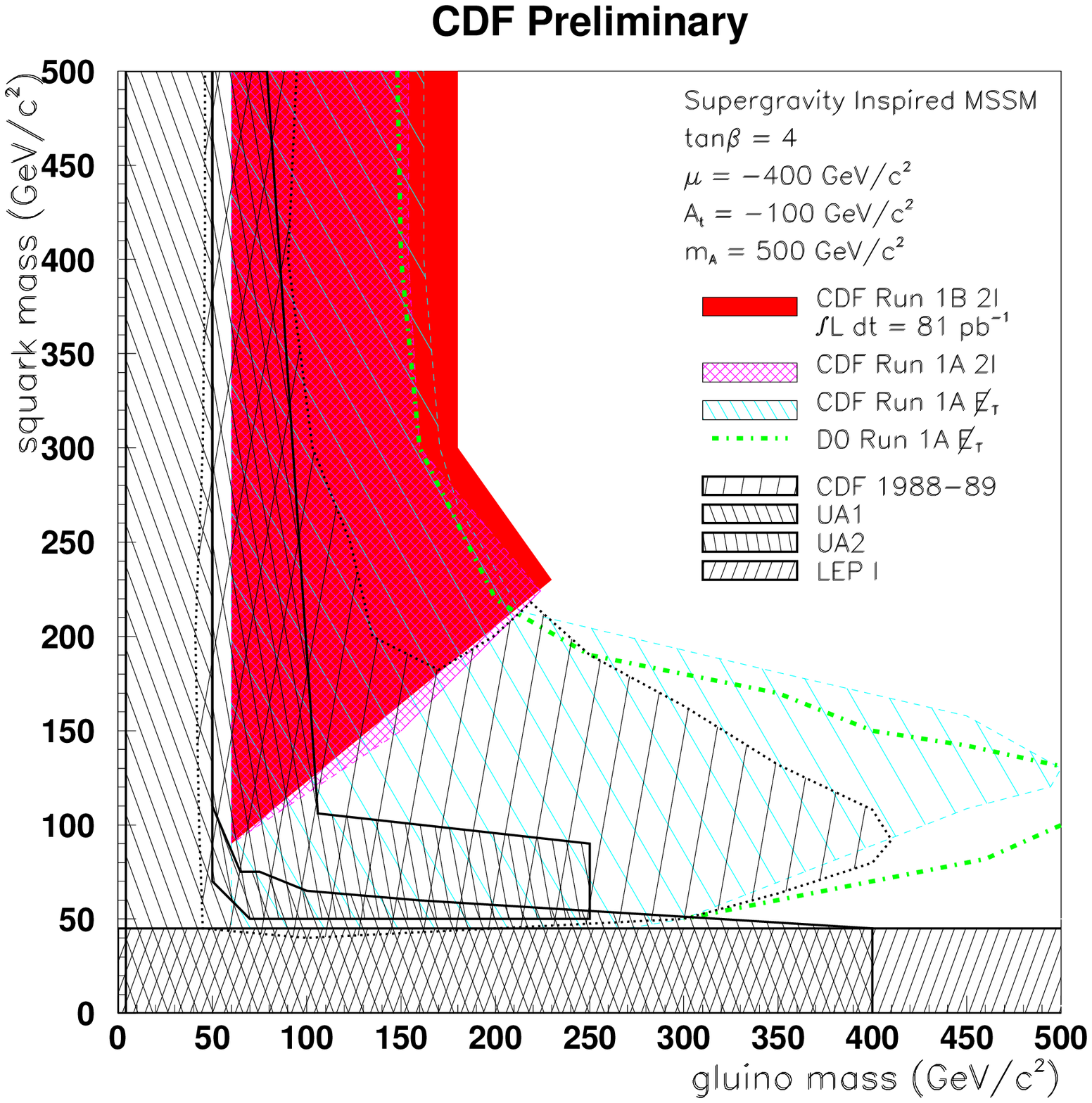,height=75mm}}
\hbox{\psfig{figure=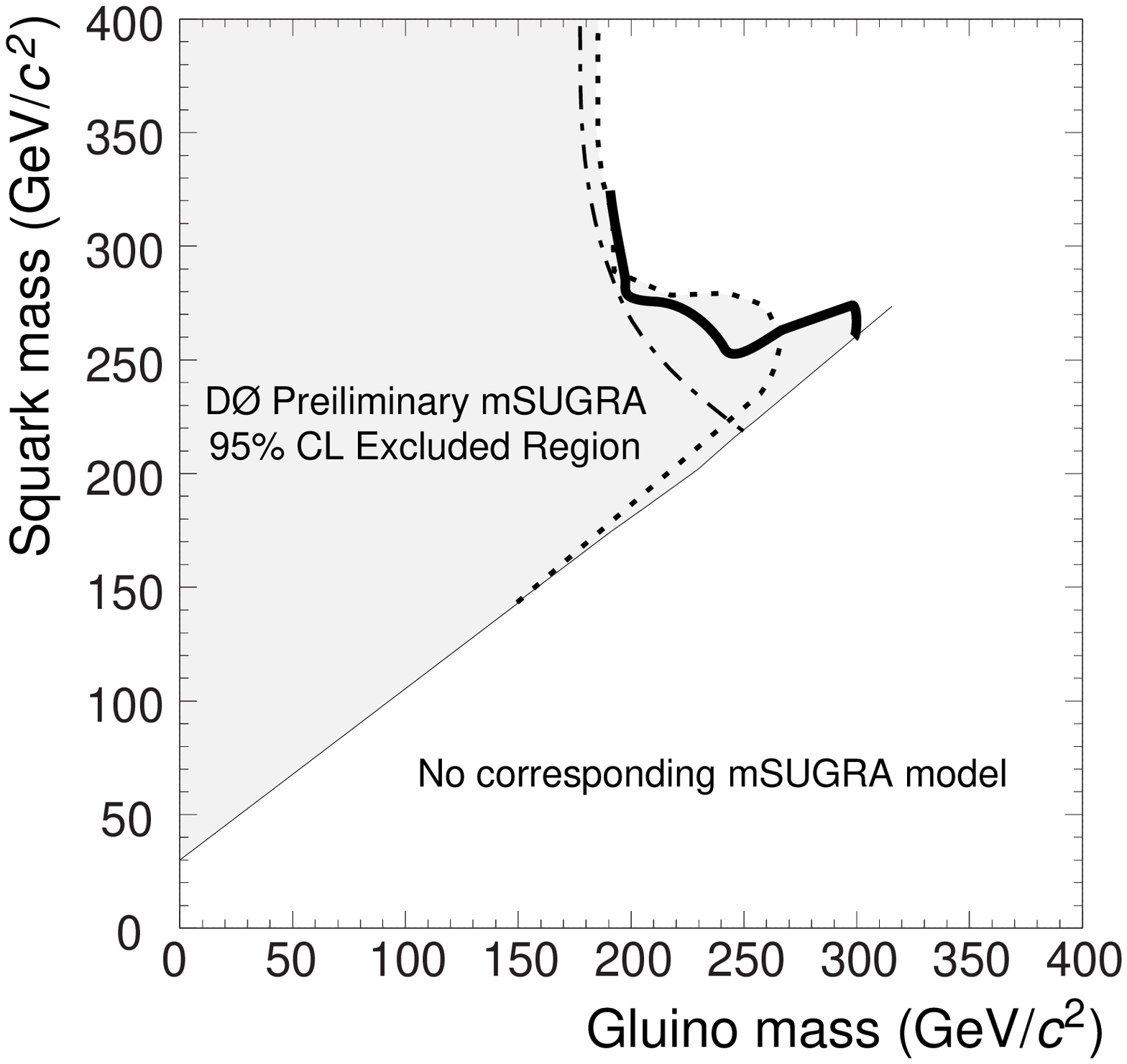,height=60mm}}
}
}
\caption{ (Left) CDF limits on the $\squark$ and $\gluino$ masses from the
2 like--sign leptons, 2 jets, and \met~search in 81 \ipb (dark shading).  
The limits were set using the ISAJET 7.06 Run I Parameter Set (RIPS)
with the indicated values.
(Right)  Excluded region from various \D0~analyses
in the $m_{\squark}-M_{\gluino}$ plane with fixed
mSUGRA parameters $\tan\beta=2$, \a0=0, and $\mu<0$. There are no mSUGRA
models in the region to the right of the diagonal thin line. 
The heavy solid line is the limit
contour of the \D0~Run Ib 3 jets and \met~analysis. 
The dashed line is the limit
contour of the \D0~Run Ib dielectron analysis. 
The dot--dashed line is the limit contour of the \D0~Run Ia
3 and 4 jets and missing transverse energy analysis shown 
only in the region with valid mSUGRA models.}
\label{fig:cdfsquarkee}
\end{figure}

The \D0~analysis ($e$ only) requires two leptons, $E_T>$15~GeV,
of any sign, while the CDF analysis ($e$ and $\mu$) requires two 
leptons $E_T>$11,5~GeV, with the same sign.
Both analyses require \met$>25$ GeV.
Figure~\ref{fig:d0squarkee} 
shows the the \D0\,\cite{D0_metdilepton_1b} results from Run Ib, 
plotted for mSUGRA models in the \m0--\mhalf~plane.
Figure~\ref{fig:cdfsquarkee} ((left), dark shading) shows 
the CDF\,\cite{CDF_metdilepton_1b,CDF_metdilepton_1a}
result plotted in the $M_{\gluino}-m_{\squark}$ plane for a
RIPS model, and (right) a mapping of the \D0~mSUGRA results from
Fig.~\ref{fig:d0squarkee} (left) into the same coordinates.
The CDF limit is based on NLO cross sections,\cite{squarknlo} and
the \D0~limit on LO cross sections.
The \D0~limits on \m0~and \mhalf~are calculated including contributions from
the production of  all sparticles (for instance, associated production of
$\zinog$'s or $\winog$'s with $\squark$'s or $\gluino$'s),
while the CDF result considers only $\squark$ and  $\gluino$ production.

For $m_{\squark}\gg M_{\gluino}$ or, equivalently,  for 
$m_0 \gg m_{1/2}$, $\gluino\gluino$ pair production
is the dominant SUSY process. 
As $m_0 (m_{\squark})$ is varied with the other parameters fixed, the 
branching ratios for the 3--body $\gluino$ decays to $\winog$'s or $\zinog$'s
and jets become fairly constant, so the production rate of leptonic final
states becomes constant.
For large 
enough values of the $\gluino$ mass, the leptons easily
pass the experimental cuts,
and the experimental limit approaches a constant value asymptotically, as
can be seen in both the \D0~and CDF plots shown in 
Fig.~\ref{fig:cdfsquarkee}. 

The relation $m_{\squark}\ll M_{\gluino}$ is not possible in SUGRA, and
is treated in an {\it ad hoc} manner in RIPS.  There is no limit in this region
for either opposite-- or like--sign dilepton pairs because the large, fixed 
$\slepton$ masses (as assumed in this analysis) limit
the branching ratios to leptonic final states.  The possibility of 
like--sign dilepton pairs is further reduced because both
the $\gluino\gluino$ and $\gluino\squark$ cross sections 
(which produce like--sign leptons
because the $\gluino$ is a Majorana particle) and the $\squark\squark$
cross section (which produces like--sign leptons because the $\squark$'s
have the same charge) are small in this region.  

When $m_{\squark}\simeq M_{\gluino}$, the $\gluino\gluino$ cross section
is supplemented by the $\gluino\squark$ cross section.
Just above the diagonal line at $M_{\gluino}=m_{\squark}$
({\it i.e.} $m_{\squark}$ just larger than $M_{\gluino}$) in 
Fig.~\ref{fig:cdfsquarkee} 
there are ``noses'' in the limit plots, with the limit becoming stronger close
to the diagonal. 

The limits in Fig.~\ref{fig:cdfsquarkee}
are for a specific choice of parameters within RIPS or mSUGRA.
If $\mu$, $A_t$ and \tanb~are varied, the branching ratios
into $\winog$'s or $\zinog$'s can vary strongly.
The sensitive dependence on the parameters can be 
seen within mSUGRA from the \D0~limits in 
Fig.~\ref{fig:d0squarkee}.
The dip in the \tanb=2 limit (left), around \m0$=70$ \gev, 
is a point where $m_{\slepton} > M_{\zinoh} > m_{\sneutrino}$ and
BR$(\zinoh\to\nu\bar\nu\zinol)\simeq 1$,
so the detection efficiency is very sensitive to the choice of high energy
parameters \m0~and \mhalf.
In Fig.~\ref{fig:d0squarkee} (right), with \tanb=6, 
the limits are severely reduced  compared to 
Fig.~\ref{fig:d0squarkee} (left), with
\tanb=2, in the region where $m_{\squark}\gg M_{\gluino}$.
For large \tanb, the mass splittings
are reduced, and the
leptons from the $\winol$ and $\zinoh$ decays are softer.
The non--trivial shape of the limit curves results from an interplay between
the cross section being larger when \m0~and \mhalf~are smaller
(sparticle masses are smaller) and the mass splittings being
smaller.
Consequently, although the dileptons+jets+\met~ signature 
is an excellent discovery 
channel with
little SM background, it is hard to set significant parameter limits
even using mSUGRA models.

From the present analyses in the \met+jets 
and dileptons+\met~channels,
some preliminary conclusions can be drawn on the $\squark$ and 
$\gluino$ masses.
These depend, however,  on the assumed SUSY parameters.
The \D0~limit on the $\gluino$ mass effectively develops a plateau at 185 GeV
for large $m_0$ and
\tanb$=2$, and at 134 GeV for \tanb$=6$.
The CDF limit on the $\gluino$ mass is 180 GeV for \tanb$=4$
and large $m_{\squark}$.
For equal $\squark$ and $\gluino$ masses,  
the \D0~mass limit for
\tanb=2 is 267~GeV.
From the CDF RIPS analyses and $m_{\squark}\simeq M_{\gluino}$,
the limit is about 220 GeV for 
$\tan\beta=4$.  
A direct comparison of all the above results is rather difficult since
\D0~and CDF have done analyses assuming different sets of MSSM parameters. 

{\bf Stop Squarks:} 
The top squark (stop) is a special case.\cite{Godbole,Sender}
The mass degeneracy in the $\stop$ sector is expected 
to be strongly broken, and, for sufficiently large mixing, the lightest
stop $\stop_1$
can be rather light, possibly
lighter than the $\winol$.
The $\stop_1$ has about a tenth the production cross 
section\,\cite{squarknlo} of a 
$t$ quark of the same mass, because the 
cross section behaves as $\beta^3$ at threshold 
(compared to $\beta$ for fermion pairs), 
where $\beta$ is the squark velocity in the rest frame of the pair,
and only half
the scalar partners are being considered.

The stop can be produced directly as $\stop\stop^*$ pairs or,
depending on the $\stop$ mass, indirectly
in decays $t\to \stop\zinol$, or
$\winog_i\to b\stop$.
Also depending on the $\stop$ mass, one of three decay modes is expected to
dominate.
If $(a)$ $m_{\tilde{t}_1} > m_{\winol}+ m_b$, then  
$\tilde t_1$ can decay into 
$b\winoop$, followed by the decay of the $\winol$.
This can look similar to the decay $t\to bW$, but with different
kinematics and branching ratios for the final state.
Instead, if $\stop_1$ is the lightest charged SUSY particle,  
it is expected to decay exclusively through a $\winog-\sbottom$
loop as  
$(b)$ $\tilde{t}_1 \to c\zinol$, which looks quite different from
SM top decays.  Finally, the $\stop$ can decay
$(c)$ $\stop\to b W \zinol$ 
or, if it is quite heavy, $\stop\to t\zinol$ 

\begin{figure}[!ht]
\center
\centerline{\psfig{figure=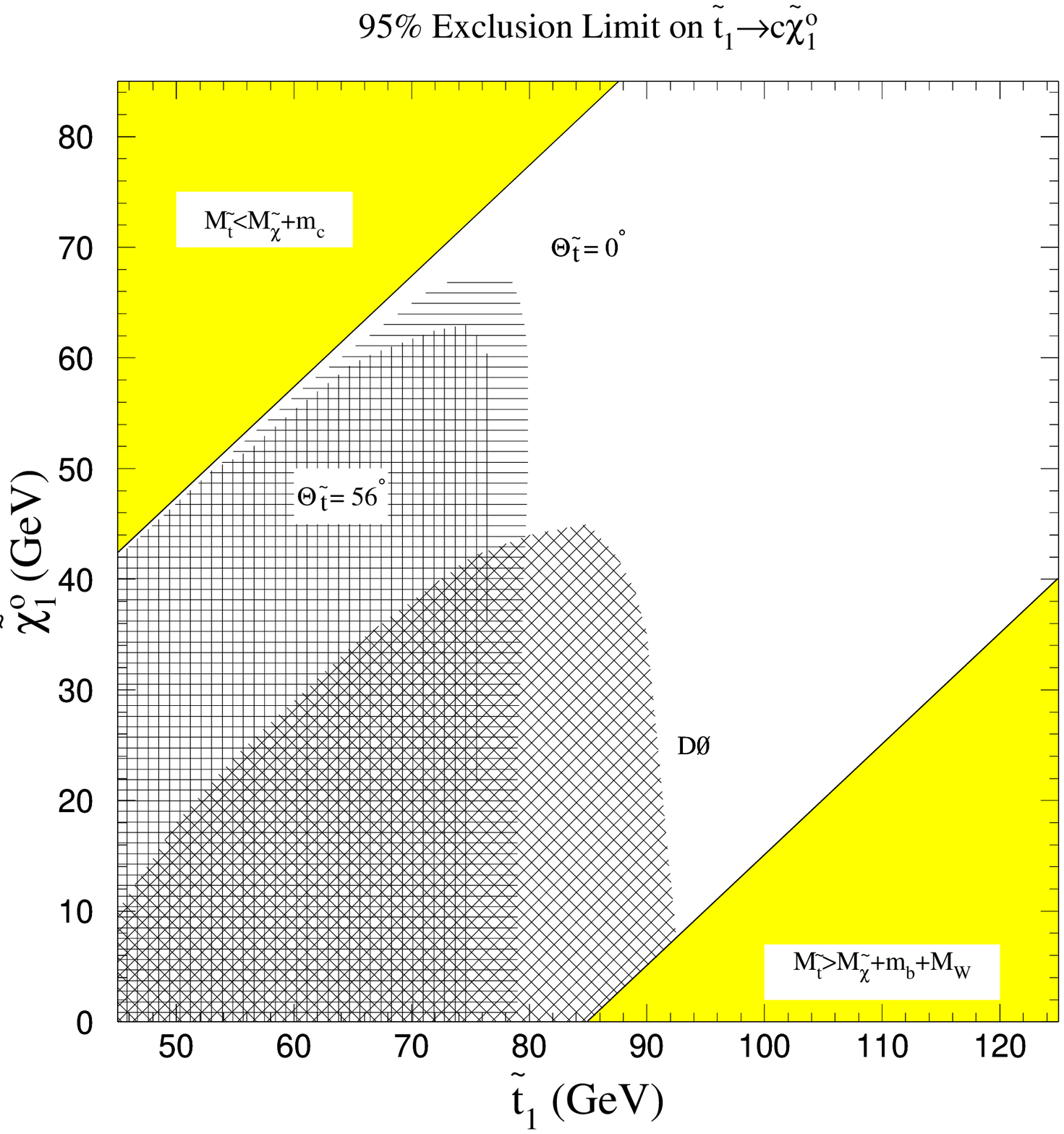,height=80mm}}
\caption{Mass limits from the \D0~search for $\stop\stop^*$ production
with the decay
$\stop\to c\zinol$ at the Tevatron.\protect\cite{D0_stopjj}   
The decay is kinematically forbidden in the two solid grey regions.
The hashed regions marked $\Theta_{\stop}$ show the 
LEP excluded regions as a function of the $\stop$ mixing angle, which
determines the strength of the $\stop$ coupling to the $Z$.
The mixing does not affect
the tree level process at hadron colliders.
}
\label{fig:d0hadstop}
\end{figure}

\D0~has searched for $\stop\stop^*$ production with
$\stop\to c\zinol$ using 7.4 \ipb~of Run Ia data.\,\cite{D0_stopjj}  
The signature used is two acollinear jets, $E_T>$30~GeV and \met$>$40~GeV.
The dijet cross section
is large, and thus this signature has large instrumental
backgrounds.   It also has 
backgrounds from vector boson production.    
The multijet backgrounds can be controlled by
requiring $\Delta \phi> 45^\circ$
between the \met~and each jet,
and that the jets not be back--to--back.  The
vector boson backgrounds are controlled by requiring that 
the two leading jets are separated by at
least $\Delta \phi>90^\circ$.    After these cuts, the dominant backgrounds
are from $W$ and $Z$ boson production and decay, with the largest being 
$W \rightarrow \tau \nu$.
The  efficiency is largest  when the stop is heavy compared to the
$\zinol$~(near the kinematic boundary for the decay $\stop_1\to bW^+\zinol$),
reaching a maximum value of only 4\%.  
The mass difference $m_{\stop}-M_{\zinol}$ determines the 
$E_T$ of the charm jet and rapidly limits
this search mode as the $c$--jets 
become too soft (see Fig. \ref{fig:d0hadstop}).
With the assumption that BR($\stop_1\to c\zinol$)=1, the predicted SUSY
final state depends only on $M_{\zinol}$
and $m_{\stop_1}$.
The result of this search is a 95\% C.L. exclusion limit on a 
region in the $M_{\zinol}-m_{\stop_1}$ plane,
shown in Fig.~\ref{fig:d0hadstop}.\,\footnote{The 
production rate has been calculated using
only LO production cross sections 
from {\tt ISAJET}.}

\begin{figure}[!ht]
\centerline{
\hbox{\hbox{\psfig{figure=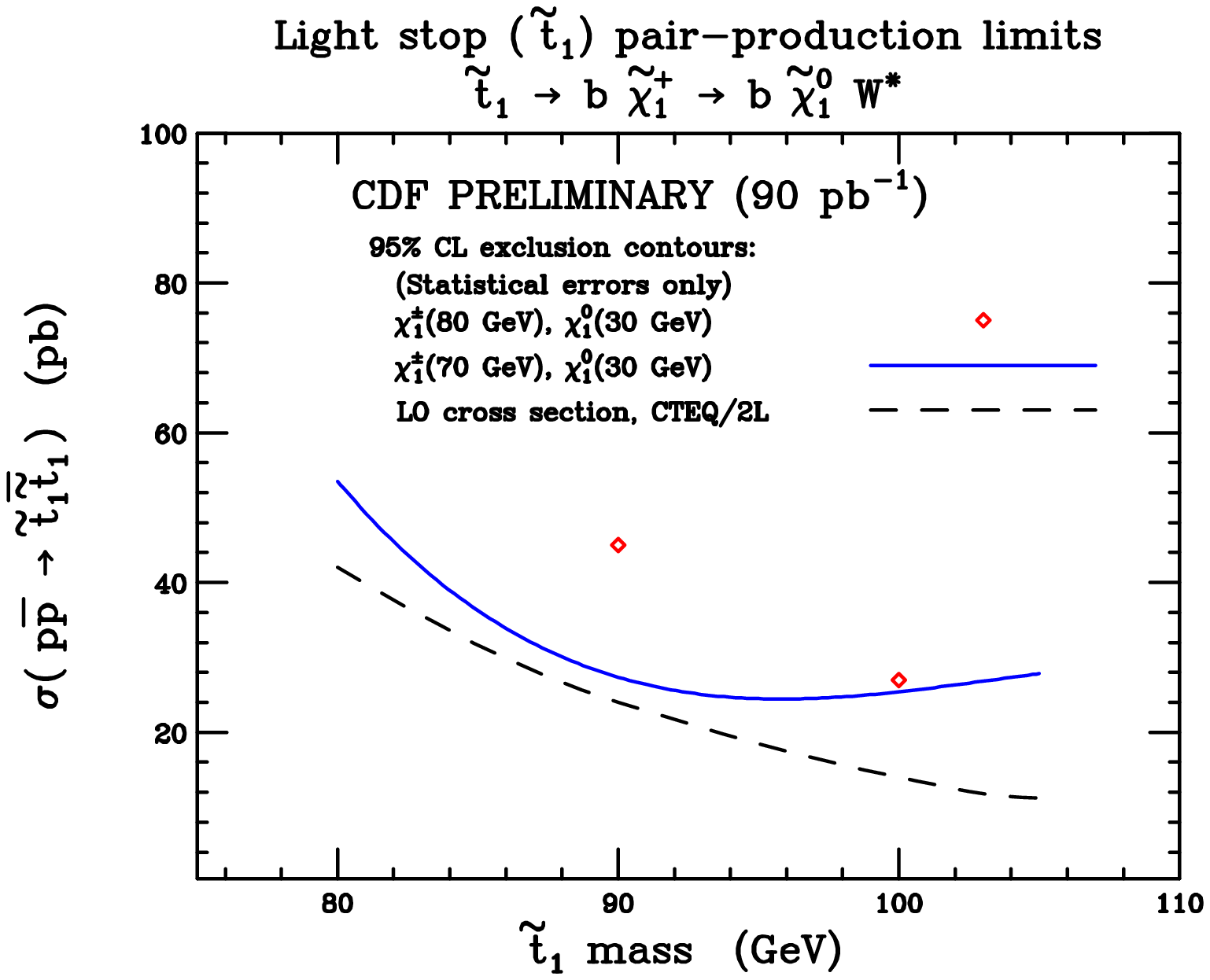,height=50mm}}
\hbox{\psfig{figure=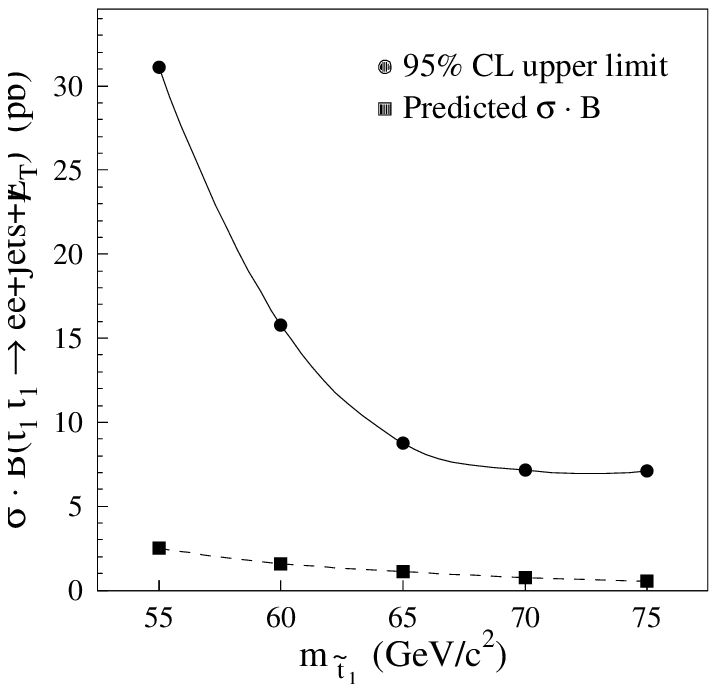,height=50mm}}}
}
\caption{(Left) The CDF cross section limit on direct production of
the top squark  using 90 \ipb~of data.  The decay mode is
$\stop\to~b\winoop(\to W^{*}\zinol)$.
One $W$ must decay semi--leptonically giving a signature of 
a lepton, \met, and jets.  
The theoretical cross section is from ISAJET 7.06.
(Right) The \D0~95\% confidence level cross section limit on  
the cross section for stop production times the branching ratio
to a final state containing 2 electrons as a function
of the mass of the $\stop$~is shown as a solid
line.\protect\cite{D0_stopee}    
The mass of the lightest
chargino is assumed to be 47~GeV.  The predicted cross section
times branching ratio from {\tt ISAJET} is also shown as a
dashed line.}
\label{fig:cdfd0stopee}
\end{figure}

CDF and \D0~have also presented results from a search for $\stop_1\stop_1^*$
production, with $\stop_1\to b\winol$.
The CDF search is in the lepton+jets channel, and
uses a shape analysis of the transverse mass
of the lepton ($E_T>$20~GeV) and~\met ($>$20~GeV).\,\cite{CDF_stop_gold}
The results of the CDF search are shown in 
Fig.~\ref{fig:cdfd0stopee} (left).  The decay $\winol\to W^*\zinol$ is
assumed using the masses $(i)$ $M_{\winol}=80$ GeV and $M_{\zinol}=30$ GeV
and $(ii)$ $M_{\winol}=70$ GeV and $M_{\zinol}=30$ GeV.\,\footnote{This analysis was done in regions of
MSSM parameter space later
excluded by LEP.  It demonstrates, however, the procedures to be followed
in performing these studies in other regions.}
Given these mass choices, there is little other parameter dependence.
Presently, the cross section limits are above the predicted cross sections 
due to the high $E_T$ cuts.

\D0~searches in the dilepton 
channel\,\cite{D0_stopee} ($E_T>$16,8~GeV) and \met$>$22~GeV.
The results are shown in Fig.~\ref{fig:cdfd0stopee} (right), assuming
$M_{\winol}=47$ GeV and $M_{\zinol}=28.5$ GeV.
A substantial background comes  from $Z\to\tau^+\tau^-$, 
again requiring a high threshold for the $E_T$ cuts, and no limit 
can be set.\,$^j$

CDF has presented another analysis using the SVX--tagged
lepton+jets sample to search for the decay $t\to\stop_1\zinol$, with
$\stop_1\to b\winol$.\cite{CDF_stop_carmine}
If one $t$ in a $t\bar t $ event decays 
$t\to bW(\to \ell\nu)$ and the other $t\to\stop_1\zinol$ followed by
$\stop_1\to b\winol(\to jj\zinol)$
or $t\to bW(\to jj)$ and  $t\to\stop_1\zinol$ followed by
$\stop_1\to b\winol(\to \ell\nu\zinol)$,
the signature is $b\bar b\ell\nu jj$+\met, 
the same as in
the SM, but where the \met~includes the momentum of the $\zinol$.
The cuts, lepton $E_T>$20~GeV, \met$>$45~GeV, and one SVX $b$--tag,
are optimized for
acceptance of the SUSY decay and rejection of $W$+jets background.
A likelihood function is computed for each event
reflecting the probability that the jets with 
the $2^{\rm nd}$ and $3^{\rm rd}$ 
highest $E_T$ in the event are consistent with the stiffer 
SM distribution (as compared to the SUSY distribution).
The distribution of this likelihood function shows a significant 
separation of these two hypotheses.
After applying the cuts, 9 events remain, 
all of which fall outside of the SUSY signal region.
For $\stop$ masses between 80 and 150 GeV and $\winog$ masses between
50 and 135 GeV, a BR($t\to \stop_1\zinol$)=50\% is excluded at the
95\% C.L., provided that $M_{\zinol}=20$ GeV.\,$^j$
Because $M_{\zinol}$ is fixed in this manner, it is not related to
$M_{\winol}$ as in SUGRA.

\subsection{Sleptons}
At hadron colliders, $\slepton$'s and
$\sneutrino$'s can only be directly produced through their
electroweak couplings to the $\gamma, Z$ and $W$ bosons.
The production cross sections are at most a few tens or hundreds
of fb at the Tevatron,\,\cite{baer} 
and the physics and instrumental backgrounds
are numerous.
So far neither collaboration has presented results on searches 
for sleptons in
the SUGRA or RIPS framework (we describe limits in gauge--mediated 
\sb models later).    

A (stable) charged slepton is not a viable LSP candidate, so the
decays $\tilde{\ell}_{L,R}^{\pm} \to \ell^{\pm} \tilde{\chi}^{0}_i $
or $\tilde{\ell}^{\pm}_L \to \nu\winog_i$ are expected.  A
$\sneutrino$, instead, can be the LSP, or it can decay invisibly
$\sneutrino\to\nu\zinol$, or visibly $\sneutrino\to\winog_i\ell^\mp$.
If $m_{\sneutrino} < m_{\tilde\ell} < M_{\zinol}$, then the decay
$\tilde{\ell} \to \ell' \nu' \sneutrino $  (or $\tilde{\ell} \to q \bar{q}
 \sneutrino $) is possible.  Promising signatures are $(i)$
$e^+e^-,\mu^+\mu^-,\tau^+\tau^-$ + \met, $(ii)$
$e\mu,e\tau,\mu\tau$ + \met, and $(iii)$ $e,\mu$ or $\tau$+jets +
\met~(or jets + \met).  
Although charged slepton production can lead to charged leptons
in the final state, there is no guarantee.

\subsection{Charged Higgs Bosons} 
\label{section:chargedhiggsresults}
Even though Higgs bosons are not sparticles,  the discovery of 
one or more could be considered {\it indirect} evidence
for SUSY.
If it is light enough, the
$H^\pm$ can be produced 
in the decay $t\to bH^+$.\,\cite{Roy}
The branching fraction for this decay depends on the 
$H^\pm$ mass and \tanb,  
and
is larger than 50\% for \tanb~less than approximately 0.7 or greater than
approximately 50, but very small or 
large values of \tanb~are theoretically disfavored.
In general, at reasonably small values of $\tan\beta$, 
$H^+\to c\bar s$; at large $\tan\beta$, 
$H^+\to \tau^+\nu_\tau$.

CDF has searched for the decay $t\to bH^+$ using both
direct\,\cite{CDF_ch_higgs_jessop,CDF_ch_higgs_rutgers} and
indirect\,\cite{CDF_trilepton_1ab} methods. 
Direct searches look for an excess
over SM expectations of events with $\tau$ leptons from the
decay $H^+\to \tau^+\nu_\tau$ (dominant for large $\tan\beta$). 
The signature for hadronically decaying $\tau$'s is  a narrow jet associated
with one or three tracks  with no other tracks nearby.
Indirect
searches are ``disappearance'' experiments, relying on the fact
that decays $t\to bH^+$ will deplete the SM decays
$t \to bW$, decreasing the number of events in the dilepton and 
lepton+jets channels.  

\begin{figure}[!ht]
\centerline{\hbox{\hbox{\psfig{figure=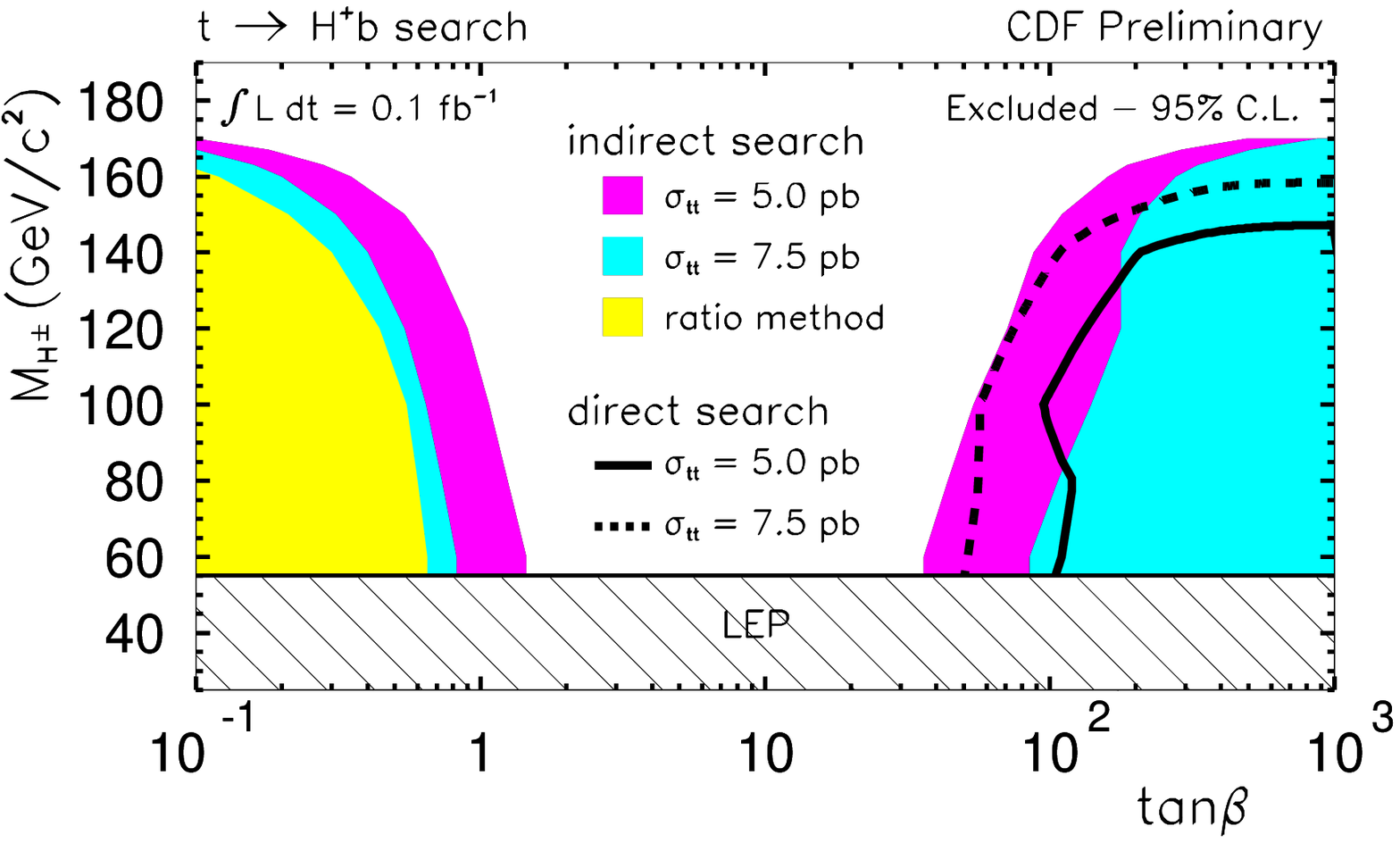,height=60mm}}
\hbox{\psfig{figure=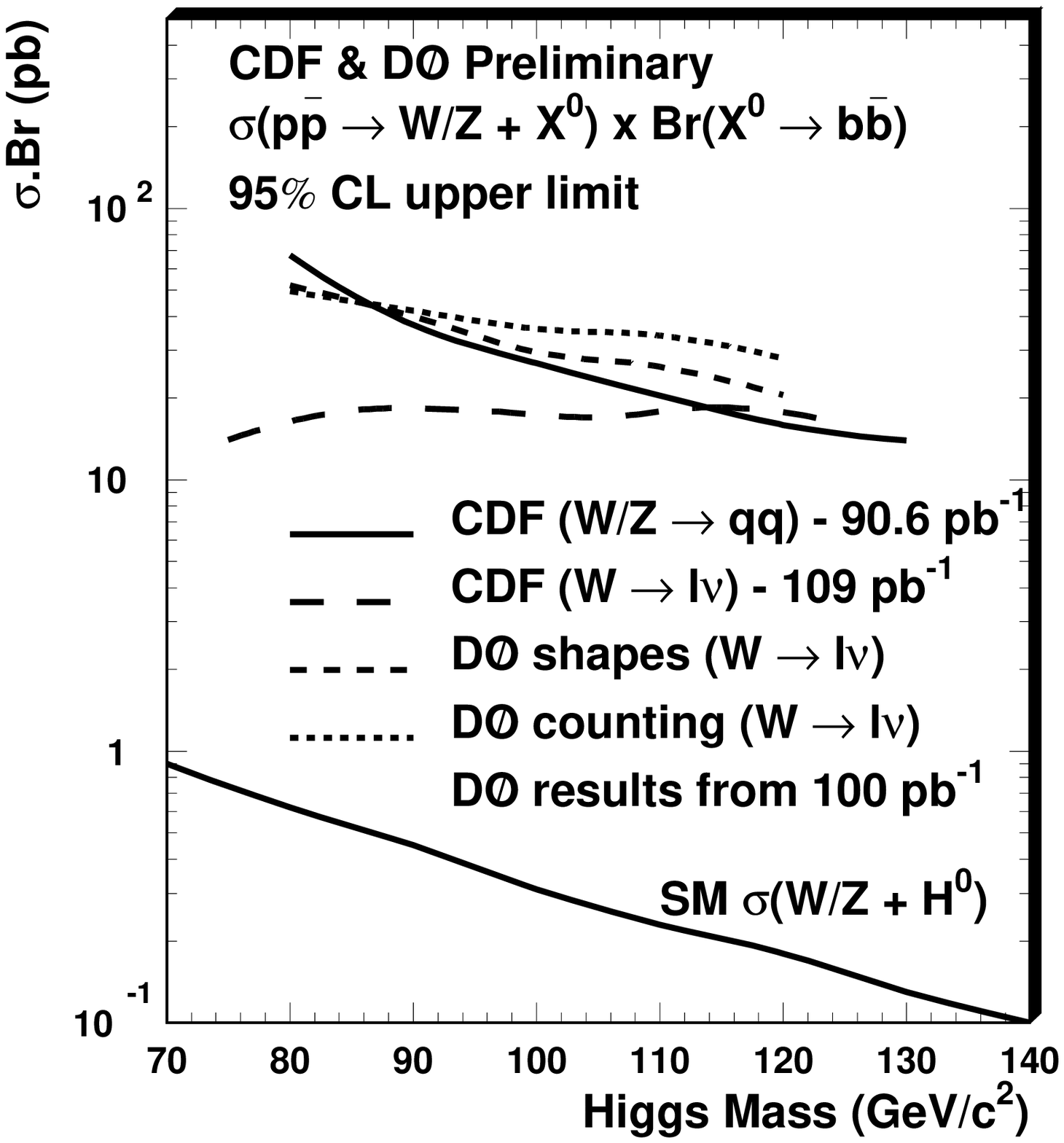,height=60mm}}}}
\caption{(Left) Exclusion space for the CDF searches for
decays $t\to bH^+$ in $t\bar{t}$ events.  
The region excluded with solid lines at high
$\tan\beta$ is from a direct search for events where one or both 
$t$'s in a $t\bar t$ event decay to $bH^+(\to\tau^+\nu)$
and information from the SM channels is ignored.
The shaded regions are from the indirect searches.
For the regions labeled $\sigma_{t\bar{t}}=5.0$ and 7.5~pb,
$\sigma_{t\bar t}$ is assumed and points are excluded if
the predicted SUSY decays have depleted the SM channels to an extent that
they are inconsistent with the data.
The ``ratio method'' is an indirect method
comparing the ratio of lepton+jets to 
dilepton events and no $\sigma_{t\bar t}$ is assumed.
(Right)
Limits from CDF and \D0~for the associated production
of $Wh$ or $Zh$.
The CDF limits are shown for the final
states of $\ell\nu$\bbbar~and $jj$\bbbar, and the \D0~limit is
for the final state $\ell\nu$\bbbar.  The limit is set using
a simple counting method and by fitting the \bbbar~spectrum (``shapes'').}
\label{figure:nhiggslimits}
\label{figure:chiggslimits}
\end{figure}

The selection criteria for the CDF direct search 
are either a single $\tau$, $E_T>$20~GeV, \met$>$30~GeV and a 
SVX $b$-tagged jet, or two $\tau$'s, with $E_T>$30~GeV.
Values of $m_{H^\pm}$ and $\tan\beta$
can be excluded based on
two methods: either the model is inconsistent
with the observation of of $\tau$'s, 
(Fig.~\ref{figure:chiggslimits} (left)), 
or the model is inconsistent with the combination of the 
number of $\tau$ events and the number of lepton+jets events.
The second method has the advantage that a $t\bar{t}$ cross section
$\sigma_{t\bar t}$ 
does not need to be assumed; 
details are presented elsewhere.\,\cite{CDF_ch_higgs_rutgers}
The reader should be aware that there are subtleties in analyses that 
assume cross sections.\,\cite{chapter_long_version}

A direct search at small \tanb~is difficult since the 
$H^\pm$ would decay into two jets.
However, the indirect method can be applied to both small and large 
\tanb~searches.
If a choice of $\sigma_{t\bar t}$, $m_{H^\pm}$, and $\tan\beta$
predicts a number of dilepton and lepton+jets events that
is inconsistent with the observations, then that set of values
is excluded
(Fig.~\ref{figure:chiggslimits} (left)).
The area in Fig.~\ref{figure:chiggslimits} (left) labeled 
``ratio method'' is the exclusion 
region for an indirect search that does not make an assumption for
$\sigma_{t\bar{t}}$.  
The limit is set by comparing the ratio of
dilepton to lepton+jets
events, since $t\to bH^+$ decays
always
decreases this ratio, regardless of $\sigma_{t\bar{t}}$.

Recent studies have shown that quantum SUSY effects 
(SUSY QCD and electroweak radiative corrections) to the decay mode 
$t\rightarrow bH^+$ (with subsequent decays into $\tau$'s) 
may be important and should be considered 
in future analyses.\,\cite{ChHiggscorr}

\subsection{Neutral Higgs Bosons }
\label{section:neutralhiggsresults}

The lightest CP--even Higgs boson $h$ can be produced at the
Tevatron in the channels $Wh$ or $Zh$.\,\cite{higgssim} 
These 
channels are relevant for large 
values of $M_A$ (the SM limit) or for small $M_A$ and small   
$\tan \beta$. 
The heavier Higgs $H$ could become
relevant for searches at an upgraded Tevatron through $ZH$, $WH$ 
production,
in some restricted region of parameter space, complementary to 
the one relevant for the $h$ searches.
In addition, the enhancement of the $b$ Yukawa coupling for
large \tanb~can enhance 
$hb\bar{b}$, $Ab\bar{b}$, and $Hb\bar{b}$ production.\,\cite{Gunionhbb}

Both collaborations have searched for 
$q\bar{q}'\rightarrow W^*\rightarrow 
W(\rightarrow e\nu,\mu\nu)h(\rightarrow b\bar{b})$.
\D0~has searched in 100~\ipb~of data using
a data sample containing a lepton ($E_T>$20~GeV), \met~($>$20~GeV) 
and two jets.\cite{d0_neu_higgs}
One of the jets must have a muon associated with it 
for $b$--tagging.
Twenty--seven events pass the selection criteria; $25.5\pm 3$ events
are expected from $Wjj$ and $t\bar{t}$.  The limits shown in 
Fig.~\ref{figure:nhiggslimits} (right) are set by a 
simple event--counting method and by 
fitting the $b\bar{b}$ dijet mass spectrum.

CDF has recently completed a similar search  for the same decay mode using
109 \ipb~of data.\cite{CDF_neu_higgs_weiming} 
Leptons must have $E_T>$25 ($e$) or 20~GeV ($\mu$)
and the event must have \met$>$25 ($e$) or 20~GeV ($\mu$)
and one SVX $b$--tag.  These events 
are split into single--tagged (one SVX tag) and
double--tagged samples (two SVX tags or one SVX and one lepton 
($e$ or $\mu$) tag). 
The 36 (6) single--tagged (double--tagged) events  are consistent with the 
$30\pm 5$ ($3.0\pm 0.6$) expected from SM $W$+jets and $t\bar{t}$.
Both the single-- and double--tagged dijet mass distributions
are fit simultaneously to set the limits shown in 
Fig.~\ref{figure:nhiggslimits} (right).

The process $q\bar q\to Z^*\to Zh$ occurs at a comparable rate to the
$W^*$ process.  CDF has searched for both processes
assuming $W/Z\to jj$.\cite{CDF_neu_higgs_valls}  
All events must have 4 jets with $E_T>$15~GeV and two SVX $b$--tags.
In 91 \ipb~of data, 589 events remain, consistent with the expectation 
from QCD heavy--flavor production and fake
tags.  To set limits, the $b\bar{b}$ dijet mass spectrum is fit.  
Also shown in Fig.~\ref{figure:nhiggslimits} (right) is the SM production 
cross section 
for $Wh$ and $Zh$ as a function of the Higgs boson mass.  The present
experimental limits are roughly two orders--of--magnitude away from the
predicted cross section.  

\D0~has also searched for an $h$ 
with suppressed couplings to fermions,\cite{d0_gg_higgs}  
so that $h\to\gamma\gamma$ can be dominant (see the
first reference of \oldcite{higgssim}).  Events are
selected containing two
photons with \et $>$ 20 and 15 \gev, and
two jets with \et $>$ 20 and 15 \gev. 
No evidence of a resonance is seen in 
the mass distribution of the 2 photons, and \D0~excludes 
such a Higgs with a mass less than 81~GeV at a 95\% C.L.

\subsection{R--Parity Violation}

Allowing for \sr~in the MSSM opens a host of
possibilities at the Tevatron.\,\cite{RPV}   
The possible excess of HERA events at large $Q^2$ has triggered interest
in studying the consequences of the interaction of a light $\squark$ 
(preferably a $\stop$ or $\tilde c$) with an electron and a $d$ 
quark.\,\cite{CandR}
If the $\gluino$ were heavier than this $\squark$, then $\gluino$ pair
production at the Tevatron and the decay 
$\tilde{g}\rightarrow \bar c{\tilde c}_L$ 
through R--conserving couplings, followed by the \sr~decay 
$\tilde{c}_L\rightarrow e^+d$, would yield the signature of two electrons
and 4 jets.\,\footnote{If the above \sr~decay is allowed, then
the same \sr~coupling will induce the decays $\tilde s_L\to\nu_e d$,
$\tilde d_R\to e^- c$ and $\tilde d_R\to \nu_e s$.}

CDF has performed a search\,\cite{CDF_Rparity} considering the 
\sr~$\squark$ decays with the signature of two like--sign electrons
($E_T>$15~GeV) and two jets ($E_T>$15~GeV).
In 105 \ipb~of Run Ia and Ib data, no events remain after all cuts 
are applied.
Varying the masses of the SUSY particles does not 
alter the acceptance significantly since they are heavy enough for
the decay products
to easily pass the $E_T$ thresholds.  Because of this, the 
limit on the cross section times branching ratio is approximately constant 
at 0.19~pb.  For $m_{\tilde c_L}=200$~GeV, this excludes
$M_{\gluino} < 230$ GeV, assuming
${\rm BR}(\tilde{g}\tilde{g}\rightarrow e^\pm e^\pm X)=1/8$.

Allowing for
R--parity conserving $\squark$ decays, the decay
$\squark\to q\zinol$ is possible, where $\zinol$ is the LSP.
Since the LSP has no R--parity conserving decays kinematically accessible,
the \sr~decay
$\zinol\to c \bar{d} e^-$ or $\bar c d e^+$ occurs through a 
virtual $\tilde c$ or $\tilde d$,
while $\zinol\to d \bar{s} \nu$ or $\bar d s \bar\nu$ 
occurs through a virtual $\tilde s$ or $\tilde d$.
For the analysis, 5 $\squark$ masses are assumed to be degenerate,
and $\squark$ masses less than 210~GeV are excluded if the 
mass of the $\zinol$ is more than half of the $\squark$ mass
and the $\gluino$ is heavy. 

If R--parity is violated, and the LSP is charged ({\it e.g.} a $\stau$),
it can be long--lived and appear as a heavy stable particle.
The particle can be identified by measuring the $dE/dx$ energy loss 
as it passes through the CDF SVX and CTC detectors.  For a 
given momentum, a heavy particle has a slower velocity and hence
a greater energy loss than a relativistic particle ($\beta\simeq 1$).  
If the particle is weakly interacting 
or massive enough to kinematically suppress showering, it will penetrate
the detectors and be triggered on and reconstructed 
as a muon with too much energy loss.
A result using part of the Run I data has been
presented by CDF \cite{CDF_stables} and is now updated with the full data set.
In 90 \ipb~of inclusive muon triggers ($p_T>$30 \gev), CDF 
searches for particles with ionization consistent with
$\beta\gamma < 0.6$ and finds 12 events 
depositing more than twice the energy expected from a minimum ionizing muon.
This is consistent with the number of events expected 
from muons which overlap with other tracks to fake a large $dE/dx$ signal.

\subsection{Photon and \met~Signatures}
\label{section:introeeggm}

SUSY has so many parameters that
the full range of its allowed signatures may be hard to predict.
In April 1995, the CDF experiment recorded 
an event with a very unusual topology\,\cite{Park} which
may have SUSY interpretations.
It has four electromagnetic clusters, which pass the
typical cuts for two electrons and two photons, and \met.

There have been two main proposals for a possible 
SUSY explanation of the event: the  Gravitino LSP and 
the Higgsino LSP model (there are also non--SUSY explanations\,\cite{nonsusy}).
In gauge--mediated $\sb$ models, the 
$\gravitino$ is the LSP and
the next--to--lightest
superpartner (NLSP)
decays  into its SM partner plus the
Goldstino component of the $\gravitino$.\,\cite{Fayet}
If $\zinol$ is the NLSP, the only modification
to SUGRA phenomenology,
where all sparticles decay down to $\zinol$, is that
$\zinol$ then decays to a photon and \met. 
In particular, the CDF event can be interpreted as either
$\selectron\selectron^*$ production\cite{thomas,ellis} or
$\winoop\winoom$ production.\,\cite{ellis}
 
The Higgsino LSP model\,\cite{n2ton1gprd} involves a region of 
MSSM parameter space in which the
$\zinoh$~is photino--like and the $\zinol$~is
Higgsino--like, so the
radiative decay $\zinoh\to\gamma\zinol$~dominates 
over other $\zinoh$ decay modes.\,\cite{radiative_decay}
The event can be again interpreted as $\selectron\selectron^*$ production
or $\winoop\winoom$~production.

Both proposals also 
suggest 
other signatures that should be expected within these models.\,\cite{others}
The $\gravitino$ LSP model predicts a large number of events
with many jets, leptons, and $\gamma$'s, and the
fact than none of these other signatures
has been detected makes the above LSP $\gravitino$ explanation of the CDF 
$ee\gamma\gamma$\met~event unlikely.\,\cite{interest,baertata} 
In Higgsino LSP models, $\gamma$'s only arise from the decay of $\zinoh$,
and there is no guarantee that there will be other
substantial signals.
A logical starting place for searches is in the inclusive
two photon and \met~channel.\cite{interest}
The generic $\gamma\gamma$\met~$+X$ signature has 
no significant background from real $\gamma$'s.  The main backgrounds are 
caused by jets and electrons faking $\gamma$'s.   
The SM production of $W(\to e\nu)\gamma$ + jets 
can fake some of the signatures if the electron is misidentified as a 
$\gamma$.
These events have a \met~spectrum typical of $W$ events, peaked
at about $M_W/2\simeq$ 40 GeV, with a long tail to high \met.    
The dominant instrumental background, however, is from di--jet and 
$\gamma+$jet production, where the large production cross section
overcomes the small probability 
($\simeq 10^{-4}-10^{-3}$) that a jet fakes
a $\gamma$. 

\begin{figure}[!ht]
\centerline{
\hbox{\hbox{\psfig{figure=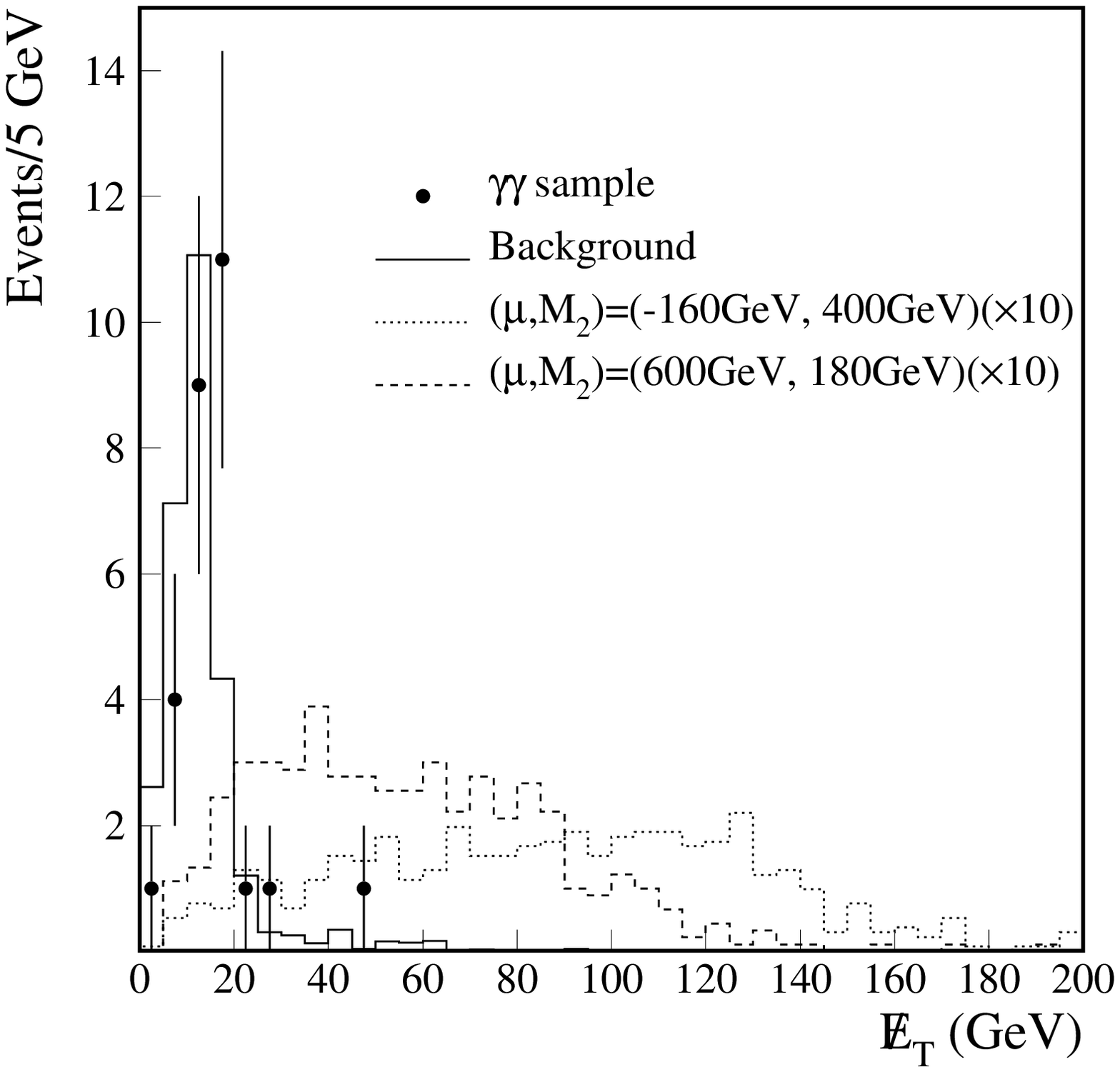,height=60mm}}
\hbox{\psfig{figure=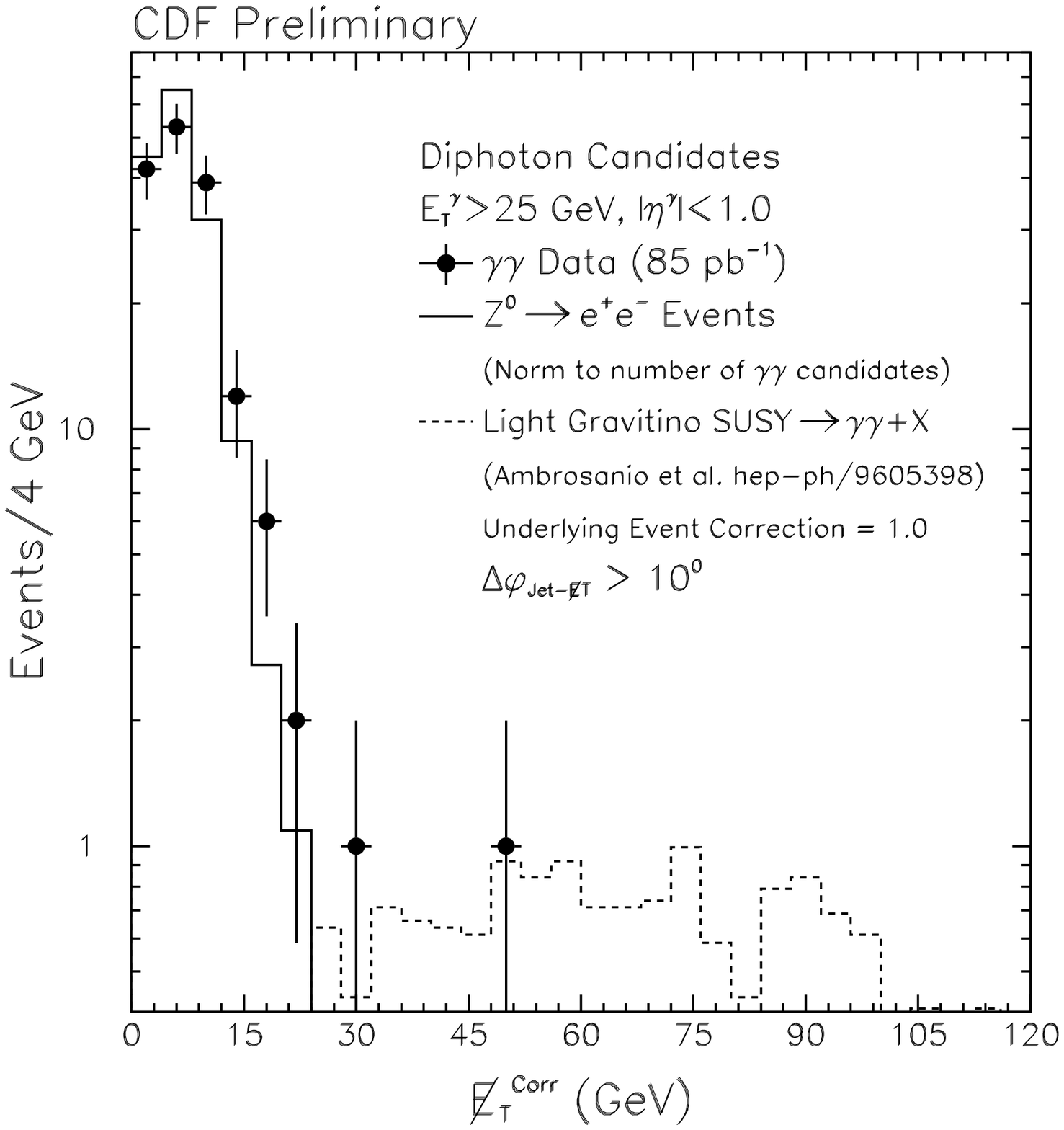,height=60mm}}
}
}
\caption{(Left) The 
\met~spectra in the \D0~search for events with 2 $\gamma$'s, one
with \et $>$ 20 \gev, 
the other with \et$>$ 12 \gev.\protect\cite{d0_diphoton}
The points are the data, the solid line is the estimated background
from di--jet events and direct $\gamma$ events.  The dotted lines
are for gaugino production within gauge--mediated models using
the parameters listed and $M_1\simeq 2 M_2$.
(Right) The CDF \met~spectrum for events with two  central
$\gamma$'s with  $E_T > 25$~GeV.  
Events which have any jet with $E_T> 10$~GeV pointing 
within 10 degrees in azimuth
of the \met~are removed.
The solid histogram shows the resolution from
the $Z\to e^+e^-$ control sample.
The dashed line shows the expected distribution from 
all SUSY production in a model~\protect\cite{interest} 
with $M_2=225$ GeV, $\mu=300$ GeV, $\tan\beta=1.5$,
and $M_{\squark}=300$ GeV.
}
\label{fig:d0gamgam}
\end{figure}

Figure~\ref{fig:d0gamgam}
shows the \met~distributions from \D0~(left) and CDF (right)
diphoton events \cite{CDF_stop_rlc,d0_diphoton} after imposing the
selection criteria.
The \D0 analysis requires two $\gamma$'s with $E_T>$20,12~GeV and 
\met$>25$~GeV, 
while CDF requires two $\gamma$'s with $E_T>$25~GeV and \met$>35$~GeV.
For the \D0~analysis, the shape of the \met~spectra agrees well 
with backgrounds containing
two electromagnetic--like clusters, where 
at least one of the two clusters fails the $\gamma$ selection criteria.  
Two events satisfy all selection criteria, with a predicted
background, dominated by jets faking $\gamma$'s, of 2.3 $\pm$0.9 events.
For the CDF analysis,
the shape of the \met~distribution is in good agreement with the
resolution of the $Z\to e^+e^-$ control sample.
The event on the tail in \met~is the ``$ee\gamma\gamma$\met'' event.  
If the source of this event is an
anomalously large $WW\gamma\gamma$ production cross section that 
yields one event in $\ell\ell\gamma\gamma$\met, 
CDF would expect dozens of events with two $\gamma$'s and 
several observed jets.  However, the jet multiplicity spectrum in
diphoton events is well--modeled by 
an exponential, and there are no diphoton events with 3 or 4 jets.


\begin{figure}[!ht]
\centerline{
\hbox{
\hbox{\psfig{figure=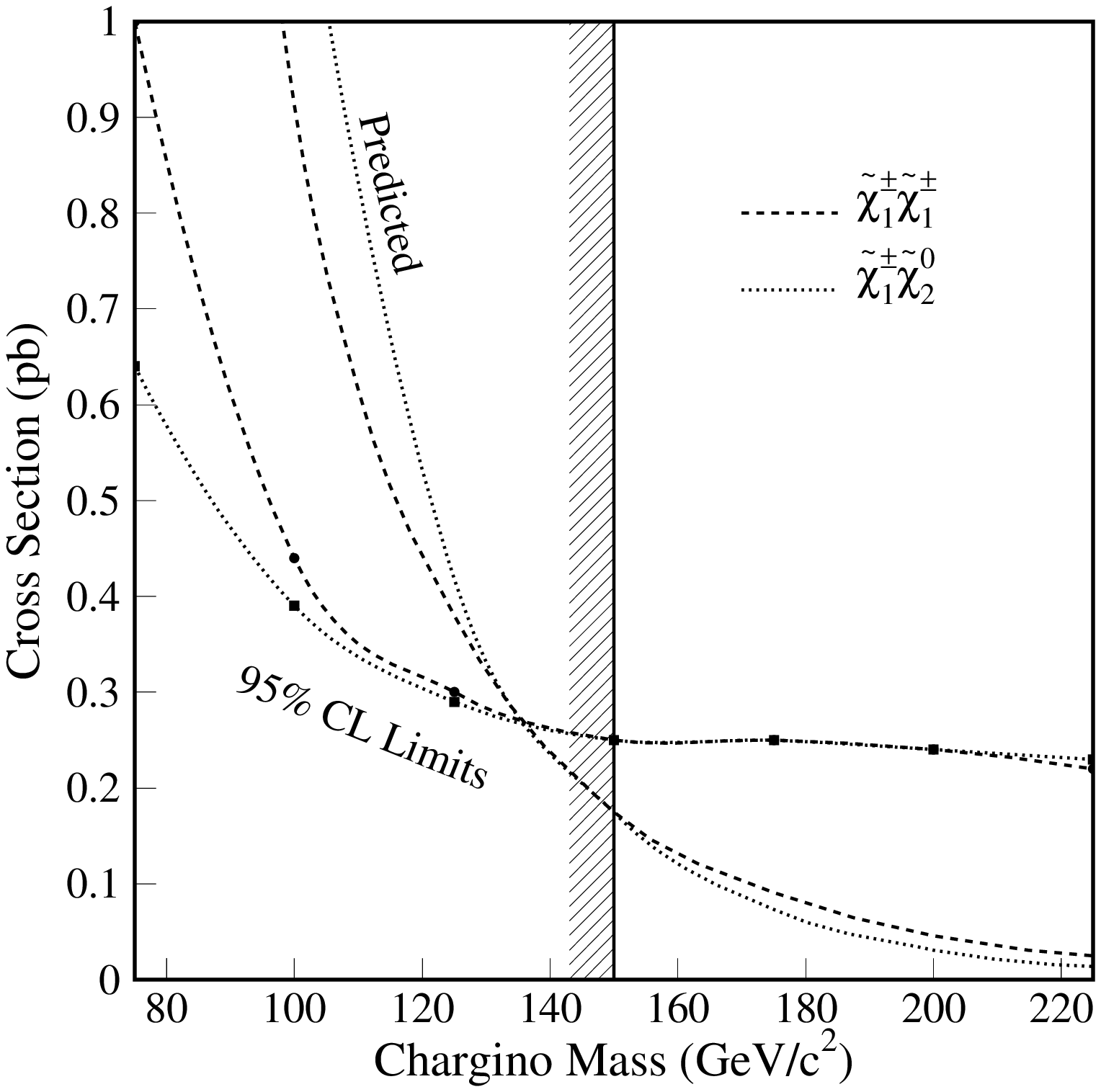,height=50mm}}
\hbox{\psfig{figure=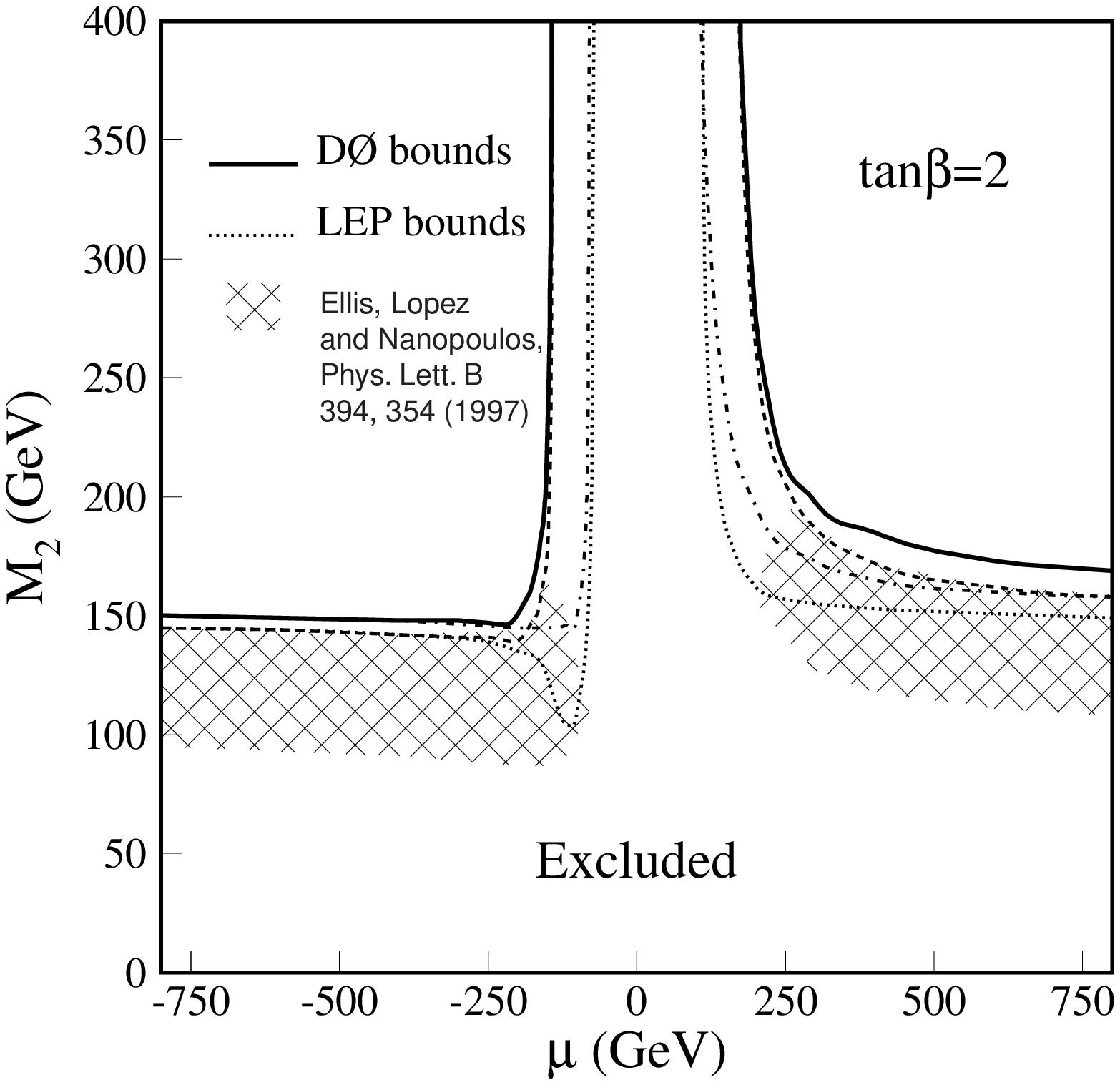,height=50mm}}
}
}
\caption{(Left) The
\D0~cross section limit on $\winol\winol$ and $\winol\zinoh$ 
production, assuming $M_{\winol}\approx 2 M_{\zinol}$ and
BR($\zinol \rightarrow \gamma \gravitino$) = 100\%.
The top dotted (dashed) curve is the cross section from 
{\tt PYTHIA} for
$\winol\zinoh$ ($\winoop\winoom$) production.   The bottom dotted
(dashed) curve is the cross section limit from the 
\D0~collaboration\,\protect\cite{d0_diphoton} 
on $\winol\zinoh$  ($\winoop\winoom$) production.
The vertical, hatched line marks the 95\% C.L. 
lower limit on the lightest chargino mass from considering
all chargino and neutralino pair production processes and all 
values of $\mu$.
(Right) The limits on the parameters $M_2$ and
$\mu$ in gauge--mediated models based on 
{\tt PYTHIA} for
\tanb=2 and $M_{\squark}$=800 \gev.\,\protect\cite{d0_diphoton}  
The hatched area is the region proposed\,\protect\cite{ellis} to explain
the CDF ee$\gamma\gamma$\met~event.
The solid line shows the \D0~bounds.
The long--dashed line shows a contour with $M_{\winol}=150$ \gev~and
the dash--dotted line shows a contour with $M_{\zinol}=75$ \gev.  
The dotted lines show an interpretation of preliminary LEP results 
at an energy of 161 GeV.}
\label{fig:d0gglimit}
\end{figure}

\D0~presents limits\,\cite{d0_diphoton}  
in the framework of the 
$\gravitino$ LSP scenario by considering $\zinog$ and $\winog$
pair production.  
Assuming $M_2 \simeq 2 M_1$ and large values of $m_{\squark}$,
the signatures are a function of only $M_2$, $\mu,$ and $\tan\beta$.
Figure~\ref{fig:d0gglimit} shows the limit
on the cross section for $\winoop\winoom$ and $\winol \zinoh$
production as a function of the $\winol$ mass when $|\mu|$ is large
and thus
the $\winol$ mass is approximately
twice the $\zinol$ mass.
The figure also shows,
more generally, the excluded region in the $M_2$--$\mu$ plane,
 along with a prediction
for the region that might explain the CDF 
$ee\gamma\gamma$\met~event as $\winoop\winoom$ production.
The latter explanation requires $100$~GeV $< M_{\winol} <150$~GeV 
with $M_{\zinol} < 0.6 M_{\winol}$ to produce one
event with a reasonable probability.
As can be seen from Fig.~\ref{fig:d0gglimit}, the cross section
limit is typically 0.24 pb for either $\winoop\winoom$ or
$\winol\zinoh$ production.  By combining all $\winog$ and
$\zinog$ pair production processes,
a $\winol$ with mass below 150~GeV is excluded.
Hence, to keep the
$\winoop\winoom$ interpretation of the $ee\gamma\gamma$\met~event,
it would be necessary to expand the analysis of Ref.~\oldcite{ellis}.

\D0~also has a limit on the cross section for
$\selectron \selectron^* \rightarrow e^- e^+  \zinoh \zinoh$,
$\sneutrino \sneutrino^* \rightarrow \nu \bar\nu  \zinoh \zinoh$, and
$\zinoh \zinoh \rightarrow \gamma\gamma\zinol\zinol$ using the
same analysis as for the $\gravitino$ LSP search.  Such
signatures might also be expected in Higgsino LSP models.
The limit on the cross section for such processes is about
0.35~pb for $M_{\zinoh}-M_{\zinol}> 30$~GeV, which is close
to the maximum cross section predicted in these models.

CDF has searched for the signature $\gamma bc\met$,
as predicted in Higgsino LSP models with a light $\stop$.\cite{kanegtm}
The data sample of 85~pb$^{-1}$ contains events with 
an isolated $\gamma$ with $E_T^\gamma>$ 25 GeV and a jet with an SVX $b$--tag.
After requiring \met$>$20~GeV,
98 events remain.\cite{CDF_stop_rlc}
The estimated background to the 98 events is $77\pm 23\pm 20$ events.
The shape is consistent with background.
About 60\% of the background is due to jets faking $\gamma$'s, 13\% to 
real $\gamma$'s and fake $b$--tags, and the remainder to SM
$\gamma b\bar{b}$ and $\gamma c\bar{c}$ production; all of these
sources require fake \met.
When the \met~cut is increased 
to 40 GeV, 2 events remain.  More than 6.4
events of anomalous production in this topology is excluded.  
The efficiency used in the limits is derived from 
a ``baseline'' model
with $M_{\zinol}=40$~GeV, $M_{\zinoh}=70$~GeV, $m_{\stop_1}=60$~GeV,
$m_{\squark}=250$~GeV, and $M_{\gluino}=225$~GeV.
The baseline model
predicts 6.65 events, so this model is excluded (at
the 95\% C.L.).  This result does not rule out 
the Higgsino \LSP~model with a light $\stop$ 
in general, only one version with a 
fairly light mass spectrum.
A more general limit can be set by 
holding the lighter sparticle masses constant and varying
the $\squark$ and $\gluino$ masses.  In this case $\squark$'s and $\gluino$'s
less than 200~GeV and 225~GeV, respectively, are excluded.

\section{Conclusions}
\label{Conclusions}
As can be seen from Table \ref{tab:summary}, there has been much
effort directed into SUSY searches at the Tevatron.  
However, given the wide range of possible experimental signatures
in the MSSM, there
is still work in progress and much to be done.  
Many Run I analyses are under way.

In Run II, two upgraded detectors at the Tevatron will collect more 
data at a higher energy of 2 TeV.  The nominal integrated luminosity
is 2 fb$^{-1}$, with a possible extension to 10 or even 30 fb$^{-1}$.
The production cross sections for heavy
sparticles will increase significantly with the higher energy, and
the $\winog$ and 
$\zinog$ searches, as well as $\squark$ and $\gluino$ searches, will cover a 
wide range of SUSY parameter space.
%
The experience gained from Run I analyses will greatly increase
the quality of the Run II searches.\,\cite{tev2000,snowmass}
New triggering capabilities will open up previously inaccessible
channels, particularly those involving $\tau$'s and heavy flavor.
Increased $b$--tagging efficiency and \met~resolution 
will enhance many analyses.
By extending Run II up to an integrated luminosity
of about 20 fb$^{-1}$ and combining search channels, the Tevatron can
perform a crucial test of the MSSM Higgs boson sector.
A factor of 20 or more data combined with improved detector capabilities
makes the next Run at the Tevatron an exciting prospect.

\section*{Acknowledgments}
The authors thank G.L. Kane for suggesting this
review, and
the following people for useful
discussions and comments: 
H. Baer,
A. Beretvas,
J. Berryhill,
B. Bevensee,
S. Blessing, 
A. Boehnlein,
D. Chakraborty,
P. Chankowski,
M. Chertok,
D. Claes,
R. Demina,
J. Done,
E. Flattum, 
C. Grosso--Pilcher,
J. Hobbs, 
M. Hohlmann,
T. Kamon,
S. Lammel,
A. Lyon, 
D. Norman, 
M. Paterno,
S. Pokorski, 
J. Qian, 
A. Savoy--Navarro,
H.C. Shankar, 
M. Spira,
D. Stuart,
B. Tannenbaum,
X. Tata,
D. Toback,
C. Wagner,
N. Whiteman,
P. Wilson
and 
P. Zerwas.

\end{document}